\documentclass[12pt]{iopart}
\usepackage{xcolor}
\usepackage{booktabs}

\usepackage{ulem}

\usepackage{graphicx}% Include figure files (loaded by jmlr2e, but explicit call is fine)

\usepackage{amssymb}
\usepackage{bm}% bold math

\usepackage{algorithm}
\usepackage{algpseudocode}

\usepackage{caption}

\usepackage{multirow}
\usepackage{array}
\usepackage{tabularx}
\usepackage{arydshln}

\begin{document}

\title[Autonomous Discovery of Ising Critical Parameters]{Autonomous Discovery of the Ising Model's Critical Parameters with Reinforcement Learning}

\author{Hai Man, Chaobo Wang, Jia-Rui Li, Yuping Tian, Shu-Gang Chen}

\address{School of Science,
Northeastern University,
Shenyang, China}

\ead{imdhios@gmail.com}

\vspace{10pt}
\begin{indented}
\item[] \today
\end{indented}

\begin{abstract}
  Traditional methods for determining critical parameters are often influenced by human factors. This research introduces a physics-inspired adaptive reinforcement learning framework that enables agents to autonomously interact with physical environments, simultaneously identifying both the critical temperature and various types of critical exponents in the Ising model with precision. Interestingly, our algorithm exhibits search behavior reminiscent of phase transitions, efficiently converging to target parameters regardless of initial conditions. Experimental results demonstrate that this method significantly outperforms traditional approaches, particularly in environments with strong perturbations. This study not only incorporates physical concepts into machine learning to enhance algorithm interpretability but also establishes a new paradigm for scientific exploration, transitioning from manual analysis to autonomous AI discovery.
\end{abstract}

\vspace{2pc}
\noindent{\it Keywords\/}: Critical Phenomena, Reinforcement Learning, Ising Model, Phase Transitions, Finite-Size Scaling

% \submitto{\JPA}
\submitto{\JSTAT}
\maketitle
% \ioptwocol
%

\section{Introduction}

Critical phenomena, as manifestations of universal behavior in physical systems near phase transitions, have long been a central focus in statistical physics. By determining key parameters such as critical temperature and critical exponents, researchers can quantify and precisely characterize this universality, revealing common behavioral patterns across different physical systems near phase transitions.

Traditionally, Finite-Size Scaling (FSS) analysis has been one of the most widely employed methods for determining critical parameters \cite{cardyFinitesizeScaling1988}. FSS leverages finite-size effects by analyzing scaling behavior of simulation data from systems of different sizes, fitting expected scaling functions to extract critical parameters. Another commonly applied approach is locating the intersection point of Binder cumulant curves across different system sizes to precisely identify the critical point \cite{binderFiniteSizeScaling1981,selkeCriticalBinderCumulant2006}. However, these traditional methods require extremely precise simulation data and are susceptible to researcher subjectivity. To mitigate this subjectivity, automated scaling analysis methods have emerged \cite{bhattacharjeeMeasureDataCollapse2001, wenzelPercolationVortices3D2008, melchertAutoScalepyProgramAutomatic2009, sorgePyfssa0762015, klockeTopologicalOrderEntanglement2022}, but these still struggle to completely avoid local optima and dependence on initial parameters.

In recent years, Machine Learning (ML) methods have provided a new paradigm for studying critical phenomena. The nonlinear learning capabilities of neural networks can reduce human bias. For example, neural network-based phase classification methods \cite{carrasquillaMachineLearningPhases2017, chngMachineLearningPhases2017, vannieuwenburgLearningPhaseTransitions2017, suchslandParameterDiagnosticsPhases2018, zhangFewshotMachineLearning2019, zhangInterpretableMachineLearning2019, liExtractingCriticalExponents2019, theveniautNeuralNetworkSetups2019, raoMachineLearningManybody2020} can learn phase characteristics to some extent autonomously and indirectly obtain phase boundaries and critical exponents. However, these supervised learning methods still depend on large amounts of labeled data. Unsupervised methods require manual analysis and interpretation of clustering results \cite{funaiThermodynamicsFeatureExtraction2020, wangUnsupervisedLearningTopological2021, rodriguesdeassiseliasGlobalExplorationPhase2022, yueIncrementalLearningPhase2022}. On the other hand, researchers have attempted to use neural networks (NNs) to directly learn the scaling functions of physical quantities to extract critical exponents \cite{yonedaNeuralNetworkApproach2023}. However, these NN-based scaling analysis methods are essentially function optimization processes that struggle to overcome local optima limitations.

To overcome the limitations of existing methods, this paper proposes a physics-inspired reinforcement learning approach that accurately locates critical parameters by training agents capable of autonomously understanding the physical mechanisms of the Ising model. Our algorithm demonstrates exploration capabilities similar to physical intuition by simulating a "phase transition" process, efficiently determining critical parameters from any initial condition. This method effectively reduces human intervention and eliminates dependence on manual data annotation and analysis. Experimental results show that even in highly perturbed environments, this method significantly outperforms existing approaches in accuracy.

The paper is structured as follows: Section II details the theoretical framework and algorithm design of our reinforcement learning method integrated with the environment; Section III describes the experimental setup and analyzes results; Section IV discusses the physical significance of our findings and their potential applications in phase transition research, while also suggesting future research directions.

% Introduction to Algorithms and Physical Environment
\section{The AMPPI Framework}
\label{sec:amppi_framwork}
\begin{figure}[htbp]
\centering
\includegraphics[width=\columnwidth]{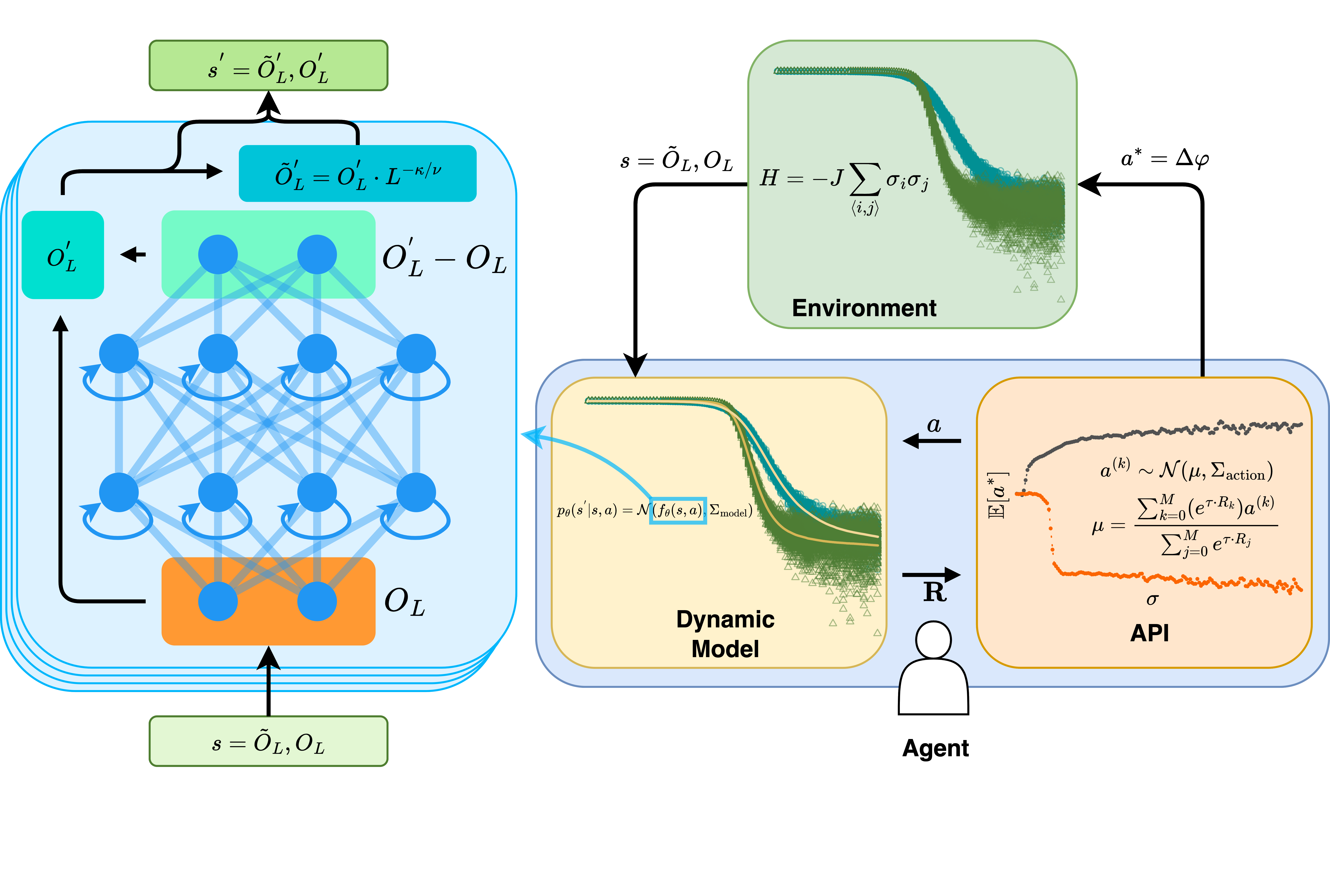}
\captionsetup{justification=raggedright, singlelinecheck=false}
\caption{Basic framework of the AMPPI algorithm. The framework includes three main components: First, the physical environment defined by the Hamiltonian; Second, the dynamic model that extracts \((O, \Delta T)\) from state-action pairs \((s, a)\) as RNN inputs to predict observable changes, then processes the results to generate new states \(s'\); Finally, the agent that samples \(K\) action sequences from \(\mathcal{N}(\mu, \Sigma_\mathrm{action})\) and generates optimal action \(a^*\) through Path Integral Control combined with REAVC mechanism.}
\label{fig:algorithm_basic_architecture}
\end{figure}

In this study, we propose an Adaptive Model Predictive Path Integral (AMPPI) algorithm framework (FIG. \ref{fig:algorithm_basic_architecture}), comprising three core components: the physical environment, a dynamic model, and an agent. The key to our method's efficiency lies in the strategic separation of computationally inexpensive internal planning from expensive physical interactions.

For the physical environment, we employ the two-dimensional Ising model with Hamiltonian:
\begin{equation}
H = -J\sum_{\langle i,j \rangle} \sigma_i \sigma_j,
\label{eq:hamiltonian}
\end{equation}
where we adopt ferromagnetic coupling \(J=1\). For a given system size \(L\), the observable physical quantities \(O_L = \{ M, E, C_v, \chi, U \}\) include magnetization, energy, specific heat, magnetic susceptibility, and Binder cumulant, along with their finite-size scaled counterparts \(\tilde{O}_L\). The environment state space is thus defined as:
\begin{equation}
\mathcal{S} = \left\{ s = \left( O_L, \tilde{O}_L \right) \mid L \in \mathbb{Z}^+ \right\}.
\label{eq:state_space}
\end{equation}

The action space includes continuous adjustments to the complete set of critical parameters, where we define \(a = \Delta\varphi\):
\begin{equation}
\mathcal{A} = \left\{ a = \Delta\varphi = (\Delta T, \Delta\beta, \Delta\gamma, \Delta\nu) \mid \Delta T, \Delta\beta, \Delta\gamma, \Delta\nu \in \mathbb{R} \right\}.
\label{eq:action_space_modified}
\end{equation}

The AMPPI algorithm operates through a two-phase framework consisting of exploration and exploitation phases to strategically balance computational efficiency with exploration effectiveness. Initially, the algorithm begins with a randomly initialized critical parameter vector \(\varphi_0 = (T_0, \beta_0, \gamma_0, \nu_0)\) sampled from predefined ranges, which generates the initial state \(s_0\) containing the physical observables through Monte Carlo simulation of the physical system.

During the exploration phase, the algorithm employs a random policy to collect state-action-state transition data \(\mathcal{D} = \{(s_t, a_t, s_{t+1})\}\) through direct interaction with the Monte Carlo-simulated environment. The collected state transition dataset \(\mathcal{D}\) is then used to train a dynamic model \(f_\theta(s,a)\), implemented as a Recurrent Neural Network (RNN) to capture subtle state changes in the system's evolution. The transition from exploration to exploitation occurs after a predetermined number of episodes ($N_{\mathrm{exp}}$), at which point the collected data enables training of a dynamic model that can effectively predict changes in the environmental states. This allows the algorithm to shift to efficient internal planning in the exploitation phase. This phase transition is distinct from the staged optimization strategy employed within the exploitation phase for multi-parameter problems.

The model parameters are optimized by minimizing the normalized mean squared error:

\begin{equation}
    \mathcal{L}(\theta) = \frac{1}{|\mathcal{D}|} \sum_{(s_t, a_t, s_{t+1}) \in \mathcal{D}} \frac{1}{2}\left\| \mathrm{Normalize}(s_{t+1}-s_t) - f_\theta(s_t, a_t) \right\|_2^2,
    \label{eq:model_loss}
\end{equation}

To enhance robustness, we employ an ensemble of RNNs with predictions combined as \(f(s_t, a_t) = \frac{1}{N} \sum_{n=1}^N f_{\theta_n}(s_t, a_t)\), where individual models are trained on different subsets of the interaction history.

The exploitation phase leverages the learned dynamic model for efficient parameter space exploration. At each step, the agent samples \(K\) action sequences of length \(H\) from a Gaussian distribution \(\mathcal{N}(\mu, \Sigma_{\mathrm{action}})\), denoted as \(\{A^{(0)}, \ldots, A^{(K-1)}\}\), where each action sequence \(A^{(k)} = \{a^{(k)}_{t}, a^{(k)}_{t+1}, \ldots, a^{(k)}_{t+H-1}\}\) represents a trajectory of parameter adjustments. The dynamic model evaluates these sequences by predicting future state evolution internally using \(\hat{s}_{t+1} = \hat{s}_t + \mathrm{Denormalization}(f_\theta(\hat{s}_t, a_t))\), avoiding expensive Monte Carlo simulations. For each sampled sequence \(A^{(k)}\), the model predicts the corresponding state trajectory and converts the predicted states to reward values, calculating the cumulative return \(R_k = \sum_{h=0}^{H-1} \gamma_d^h r(s_{t+h}, a_{t+h})^{(k)}\), where \(h\) denotes the time step index within the prediction horizon \(H\). The optimal action is then determined through Path Integral Control (PIC):

\begin{equation}
\mu_t = \sum_{k=0}^{K-1} w_k a^{(k)}_t, \quad \mathrm{where} \quad w_k = \frac{e^{\tau R_k}}{\sum_{j=0}^{K-1}e^{\tau R_j}}.
\label{eq:pic_update}
\end{equation}

The selected optimal action \(a^*_t = \mu_t\) updates the current critical parameters as \(\varphi_{t+1} = \varphi_t + a^*_t\). Only this final selected action requires physical simulation to obtain the actual next state \(s_{t+1}\) and reward \(r_t\), with the transition \((s_t, a^*_t, s_{t+1})\) stored for potential model updates.

\begin{figure}[htbp]
    \centering
    \includegraphics[width=\columnwidth]{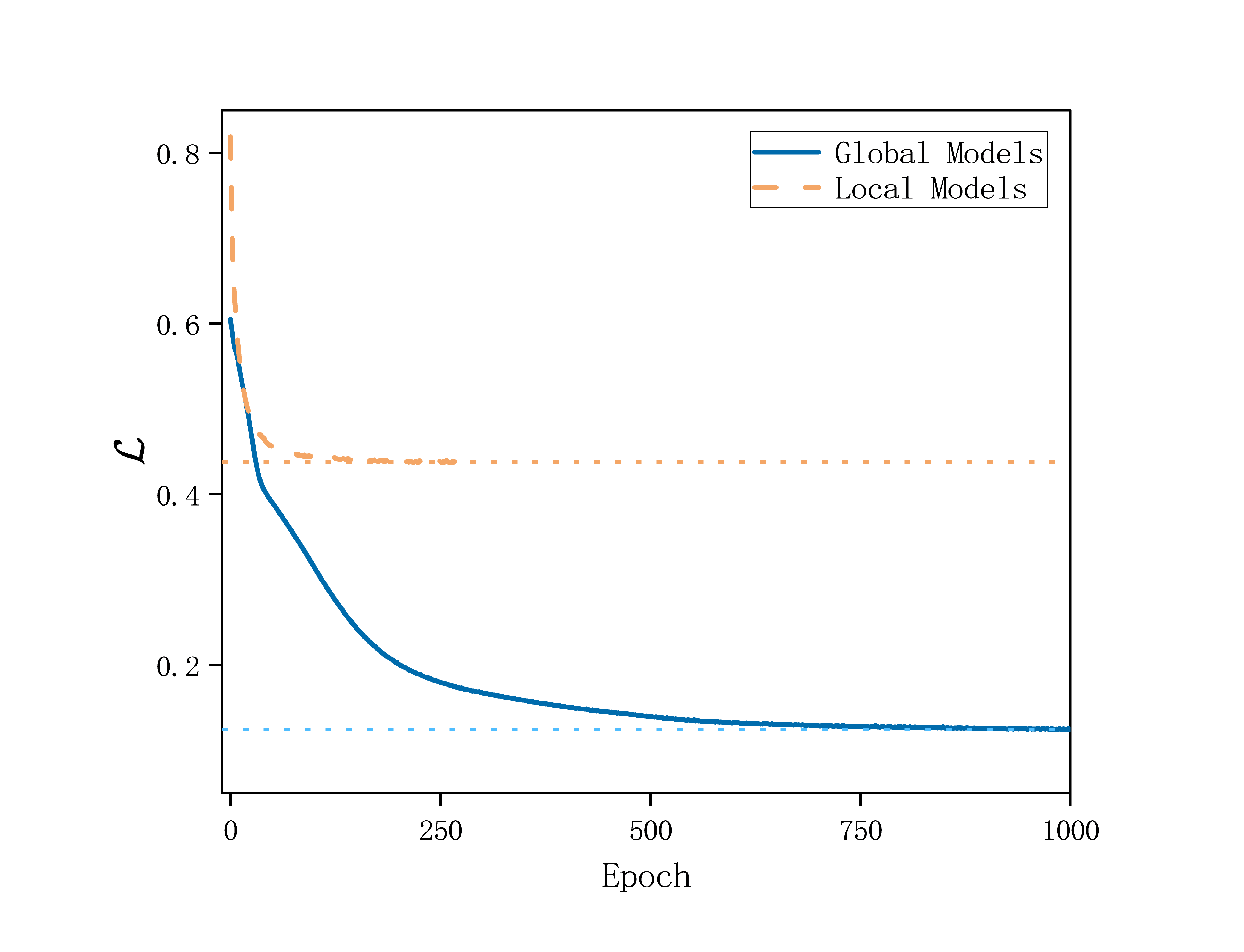}
    \captionsetup{justification=raggedright, singlelinecheck=false}
    \caption{Training convergence of two types of RNN dynamic models showing validation loss evolution for global models (solid line) and local models (dashed line).}
    \label{fig:rnn_training_process}
\end{figure}

We employ two types of dynamic models to learn from data acquired in these two phases. The global model, trained on the broader dataset collected during the exploration phase, achieves lower loss values due to smoother state transitions away from critical points. The local model, trained on data concentrated near critical temperatures during the exploitation phase, exhibits higher loss values due to inherent critical fluctuations but provides enhanced sensitivity to parameter changes in the target region, as illustrated in Figure \ref{fig:rnn_training_process}.

These complementary strengths enable effective interaction between the models during exploitation. In the exploitation phase, the global and local dynamic models interact via a threshold-based switching mechanisms driven by the Binder cumulant reward $ R(U) $ (Eq.~\ref{eq:reward_function}), as detailed in Section \ref{subsec:global_local_switching}.: global models dominate when $ R(U) < 0.5 $ for robust exploration, shifting to local models for precise refinement once $ R(U) \geq 0.5 $.

This approach fundamentally differs from traditional finite-size scaling methods such as pyfssa, which require manual extensive data collection through repeated simulations near the critical temperature. In contrast, AMPPI automatically collects physical system state change data to learn a dynamic model that can efficiently simulate the physical system, enabling efficient critical parameter analysis.

When the physical system is in a certain phase, parameter changes have minimal impact on environmental changes, leading to local optimization problems. To overcome this limitation of local optima convergence, we introduce the Reward-Error Adaptive Variance Control (REAVC) mechanism. This approach dynamically adjusts exploration variance \(\Sigma_{\mathrm{action}}\) based on reward evolution, analogous to temperature control in simulated annealing. When rewards stagnate or decrease, REAVC increases variance to promote broader exploration; when rewards improve consistently, it reduces variance for precise parameter refinement. The variance update rule is:

\begin{equation}
\Sigma_t = \xi_t \Sigma_{t-1},
\label{eq:reavc_variance_update}
\end{equation}
with the scaling factor \(\xi_t = \mathrm{clip}\left(\exp(-\beta_{RE} \Delta r) - \eta_1 e_{\mathrm{inst}} - \eta_2 e_{\mathrm{hist}}, \varepsilon_{\min}, \varepsilon_{\max}\right)\).
where \(\Delta r = r_t - r_{t-1}\) represents reward change, \(e_{\mathrm{inst}} = r_t - \alpha_{RE} \cdot r_{\mathrm{best}}\) is the instantaneous error, and \(e_{\mathrm{hist}} = \bar{r}_t - \alpha_{RE} \cdot r_{\mathrm{best}}\) is the historical error with \(\bar{r}_t\) being the exponentially weighted moving average of rewards. This mechanism, inspired by symmetry-breaking phenomena in physical systems, enables the algorithm to develop systematic search heuristics and avoid local optima effectively, with implementation details provided in Section \ref{subsubsec:reavc}.

Based on the components described above, the overall workflow of the AMPPI algorithm is summarized in Algorithm \ref{alg:amppi_main}. For a comprehensive description including detailed initialization steps and hyperparameter settings, please refer to \textbf{Algorithm \ref{alg:amppi}} in the Supplementary Material.

\begin{algorithm}[H]
    \caption{Adaptive Model Predictive Path Integral (AMPPI) Framework}
    \label{alg:amppi_main}
    \begin{algorithmic}[1]
    \State \textbf{Input:} Exploration episodes $N_{\mathrm{exp}}$, exploitation episodes $N_{\mathrm{exploit}}$, prediction horizon $H$, REAVC parameters
    \State \textbf{Output:} Optimal critical parameters $\varphi^*$
    
    \vspace{0.5em} 
    \State \textbf{Phase 1: Exploration} 
    \State Collect state transition dataset $\mathcal{D}$ using random policy $\pi_0$ for $N_{\mathrm{exp}}$ episodes
    \State Train ensemble dynamic models $f_\theta$ on $\mathcal{D}$ via Eq.~(\ref{eq:model_loss})
    
    \vspace{0.5em}
    \State \textbf{Phase 2: Exploitation}
    \State Initialize critical parameters $\varphi$ and exploration variance $\Sigma_t$
    \For{episode $= 1$ to $N_{\mathrm{exploit}}$}
        \State Reset environment to current parameters $\varphi$
        \For{step $t = 1$ to $T_{\mathrm{steps}}$}
            \State Sample $K$ action sequences of length $H$: $A^{(k)} = \{a^{(k)}_t, \dots, a^{(k)}_{t+H-1}\} \sim \mathcal{N}(\mu, \Sigma_t)$
            \State Predict state trajectories over horizon $H$ using model $f_\theta$ (Internal Planning)
            \State Calculate cumulative rewards $R_k$ for each sequence
            \State Compute optimal action $a^*_t$ (first step) using Path Integral Control (Eq.~\ref{eq:pic_update})
            \State Execute $a^*_t$ in physical environment, observe $s_{t+1}$ and $r_t$
            \State Update exploration variance $\Sigma_{t+1}$ via REAVC (Eq.~\ref{eq:reavc_variance_update})
            \State Update critical parameters $\varphi \leftarrow \varphi + a^*_t$
        \EndFor
        \State Retrain dynamic model $f_\theta$ with accumulated interaction data $\mathcal{D}$
    \EndFor
    \end{algorithmic}
\end{algorithm}

\section{Experimental Design and Performance Evaluation}

This study designed a benchmark test to evaluate algorithm performance, including: precise identification of critical temperature, simultaneous determination of critical temperature and critical exponents, and verification of the algorithm's transfer capability across different lattice structures. The following details each test protocol and results.
 
\subsection{Finding Critical Temperature}
\label{sec:critical_temperature_search}

\begin{figure}[htbp]
    \centering
    \includegraphics[width=\columnwidth]{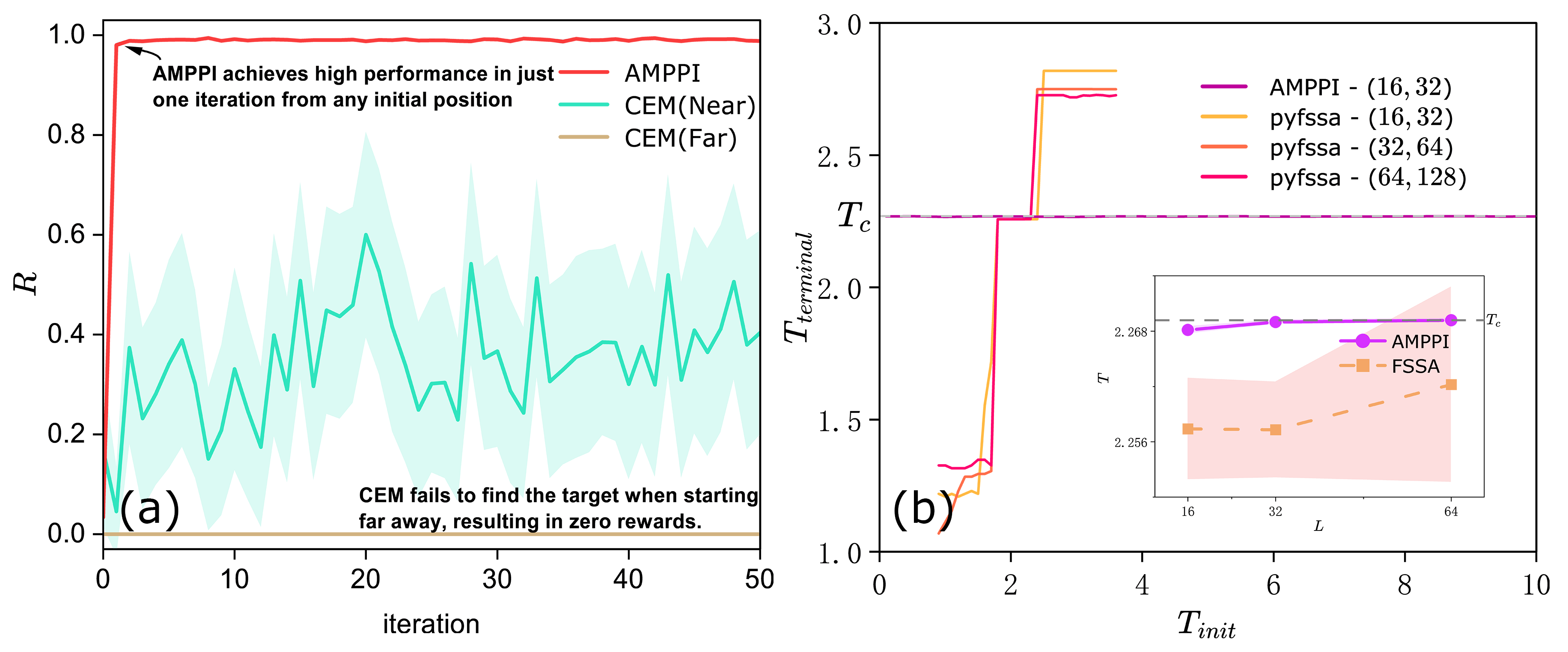}
    \captionsetup{justification=raggedright, singlelinecheck=false}
    \caption{(a) Performance curves of AMPPI and CEM algorithms in the two-dimensional square lattice Ising model (sizes 32 and 64). (b) Critical temperature determination results of AMPPI and pyfssa methods in the two-dimensional square lattice Ising model with different lattice sizes, where the horizontal axis represents initial temperature ($T_{\mathrm{init}}$), the vertical axis represents final temperature ($T_{\mathrm{terminal}}$), and the horizontal dashed line indicates the theoretical critical temperature $T_c = 2/\ln(1+\sqrt{2})$. The inset compares critical temperatures obtained by both algorithms at different lattice sizes ($L$).}
    \label{fig:learning_curve_tc}
\end{figure}

To systematically evaluate AMPPI algorithm performance, we conducted experiments with different system size combinations (16,32), (32,64), and (64,128) in the two-dimensional square lattice Ising model. To optimize computational efficiency, we used the Wolff algorithm\cite{wolffCollectiveMonteCarlo1989} for Monte Carlo simulations (see Section \ref{subsec:ising_params} for detailed environment parameters).

At each temperature point, after the system reaches equilibrium, we measure the Binder cumulant: \(U_L = 1 - \frac{\langle M^4 \rangle}{3\langle M^2 \rangle^2}\). Based on the property that Binder cumulant curves of different system sizes intersect at the critical temperature, we construct the reward function:

\begin{equation}
    R(U) = e^{-d(\mathcal{N}(U))},
    \label{eq:reward_function}
\end{equation}

Here, \(d(\mathcal{N}(U)) = \sqrt{\sum_{i=1}^{n} (\mathcal{N}(U)_{L_{i+1}} - \mathcal{N}(U)_{L_i})^2}\) quantifies the distance between normalized Binder cumulants of different system sizes. We applied Min-Max normalization to the Binder cumulant: \(\mathcal{N}(U_L) = \frac{U_L - U_{\min}}{U_{\max} - U_{\min}}\), where \(U_{\min}\) and \(U_{\max}\) are the minimum and maximum values of Binder cumulants across different system sizes under current conditions. When the system is at critical temperature, the normalized Binder cumulant curves intersect, the distance \(d(\mathcal{N}(U))\) approaches zero, and the reward value reaches its maximum of 1.

Experimental results demonstrate the exceptional performance of the AMPPI algorithm. Even a dynamic model trained with random policy sampling trajectories achieves excellent performance, as shown in FIG. \ref{fig:learning_curve_tc}(a), with an average reward value of 0.95 in the first iteration. We selected the Cross-Entropy Method (CEM) as a benchmark comparison algorithm (see Section \ref{subsec:cem_impl}). Results show significant advantages of the AMPPI algorithm: when the initial temperature is far from the critical point \(|T-T_c|>10\), the CEM algorithm fails due to its inability to determine the correct search direction; even near the critical temperature (\(|T-T_c|<2\)), its performance is severely affected by environmental perturbations, with the highest average reward value in iterations never exceeding 0.6.
 
Comparing the AMPPI algorithm with the traditional automatic finite-size scaling method pyfssa, as shown in FIG. \ref{fig:learning_curve_tc}(b), AMPPI consistently converges to near the theoretical critical temperature across a wide range of initial temperatures. In contrast, the pyfssa algorithm shows clear sensitivity to initial values: with environment sizes \((16, 32)\), its effective working range is limited to the narrow temperature interval \((1.8 \sim 2.4)\), which further narrows with increasing system size. In more distant initial temperature ranges \(0.9\sim1.7\) and \(2.5\sim3.6\), the algorithm tends to get trapped in local optima, while completely failing at initial temperatures farther from the critical point. AMPPI overcomes this limitation, eliminating the need to consider initial value selection. The inset in FIG. \ref{fig:learning_curve_tc}(b) further quantifies the precision difference between the two methods: at all tested scales, the critical temperature estimates obtained by AMPPI algorithm demonstrate significantly better precision and smaller uncertainty than pyfssa. Particularly in the large-scale system \(L=64\), AMPPI's critical temperature of \(2.2691(2)\) deviates from the theoretical value \(T_c=2.26918...\) by less than \(1 \times 10^{-4}\), improving precision by nearly two orders of magnitude and reducing uncertainty by approximately 50 times compared to pyfssa's result of \(2.2622(106)\).

\subsection{Simultaneously Finding Critical Temperature and Critical Exponents}

In critical phenomena research, traditional automated analysis methods are often limited by observable selection, typically determining only one type of critical parameter at a time. In contrast, environment-interaction learning methods overcome this limitation by processing multiple physical observables simultaneously. This section demonstrates how the AMPPI algorithm precisely determines critical temperature and related critical exponents (\(T_c\), \(\beta\), \(\gamma\), \(\nu\)) simultaneously.

Based on the previous environment settings, we apply finite-size scaling theory to analyze critical behavior. Near the critical point, the scaling relations for magnetization \(M\) and magnetic susceptibility \(\chi\) are:
\begin{eqnarray}
    M(L,t) &\sim& L^{-\beta/\nu} \tilde{M}(tL^{1/\nu}), \\
    \chi(L,t) &\sim& L^{\gamma/\nu} \tilde{\chi}(tL^{1/\nu}),
\end{eqnarray}
where \(t=(T-T_c)/T_c\) represents reduced temperature, and \(\tilde{M}\) and \(\tilde{\chi}\) are scaling functions. When the system is at the critical point (\(t=0\)) and all critical exponents are accurate, scaling function curves for different system sizes will perfectly overlap. This leads us to construct the following global reward function:
\begin{equation}
    R_{\mathrm{global}} = \frac{1}{3}\left[e^{-d(\mathcal{N}(U))} + e^{-d(\mathcal{N}(\tilde{M}))} + e^{-d(\mathcal{N}(\tilde{\chi}))}\right],
    \label{eq:multi_reward_function_global}
\end{equation}
where \(d(\mathcal{N}(\cdot))\) quantifies the distance between normalized observables of different system sizes, as defined in the previous section.

However, \(R_{\mathrm{global}}\) alone faces an inherent non-uniqueness problem. The collapse quality for \(\tilde{M}\) and \(\tilde{\chi}\) depends only on the exponent ratios \(\beta/\nu\) and \(\gamma/\nu\), providing no information about the absolute magnitudes of individual exponents. Specifically, at $T_c$ ($t=0$), the finite-size scaling $O_L = L^{\kappa/\nu} \tilde{O}(t L^{1/\nu})$ yields only ratios $\kappa/\nu$ (e.g., $O_{L_1}/O_{L_2} = (L_1/L_2)^{\kappa/\nu}$), preventing independent $\kappa$ and $\nu$ determination without off-critical ($t \neq 0$) data. Consequently, while the reward can guide the agent to correct proportions between exponents, it cannot uniquely determine \(\beta\), \(\gamma\), and \(\nu\).

To resolve this non-uniqueness, we construct an auxiliary reward function \(R_{\mathrm{aux}}\) using data near the critical temperature that specifically targets the correlation length exponent \(\nu\). This approach leverages the scaling property of the Binder cumulant: \(U_L(T) = f_U((T - T_c)L^{1/\nu})\). The auxiliary mechanism performs a "what-if" analysis by selecting a reference system \((L_j, T_j)\) and calculating equivalent temperatures \(T_k = T_c + (T_j - T_c)(L_j/L_k)^{1/\nu}\) for other system sizes \(L_k\). The agent uses its learned dynamics model to predict Binder cumulant values at these equivalent temperatures, with the auxiliary reward defined as:
\begin{equation}
R_{\mathrm{aux}}(\nu) = e^{-d_{\mathrm{aux}}(\mathcal{N}(\{U'\}))},
\label{eq:R_aux_definition}
\end{equation}
where \(d_{\mathrm{aux}}\) measures the distance between predicted Binder cumulant values. This construction provides an unambiguous learning signal sensitive only to \(\nu\), with detailed derivation provided in Section \ref{subsubsec:aux_nu_search}.

The complete optimization is guided by a composite reward:
\begin{equation}
R = \frac{1}{2}\left[R_{\mathrm{global}} + R_{\mathrm{aux}}\right],
\label{eq:composite_reward}
\end{equation} 

To ensure stable convergence, we employ a staged optimization strategy: initially focusing on \(T_c\) identification through \(R_{\mathrm{global}}\) until \(e^{-d(\mathcal{N}(U))} \geq 0.9\), then activating joint optimization of all parameters.

\begin{figure}[htbp]
    \centering
    \includegraphics[width=\columnwidth]{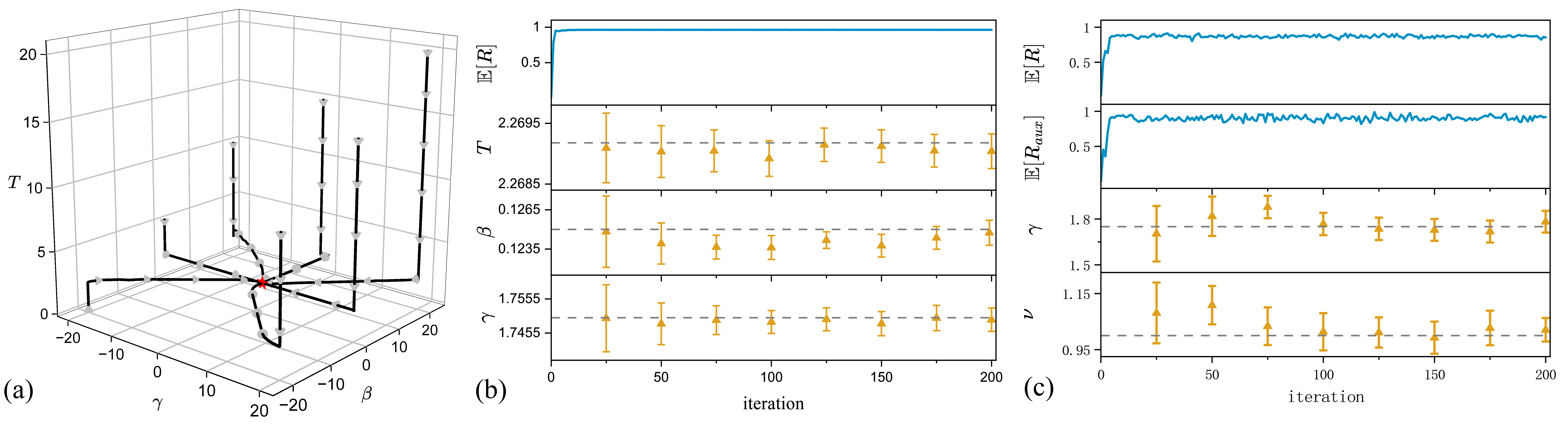}
    \captionsetup{justification=raggedright, singlelinecheck=false}
    \caption{Critical parameter optimization using the AMPPI algorithm in the two-dimensional square lattice Ising model (\(L=(32, 64)\)). (a-b) Three-parameter optimization with fixed \(\nu=1\): (a) parameter search trajectories in \((T_c, \beta, \gamma)\) space, (b) evolution of parameters and reward values over iterations. (c) Evolution of critical exponents \((\gamma, \nu)\), auxiliary reward \(R_{\mathrm{aux}}\), and composite reward over iterations at fixed critical temperature, with \(R_{\mathrm{global}} = \frac{1}{2}\left[e^{-d(\mathcal{N}(U))} + e^{-d(\mathcal{N}(\tilde{\chi}))}\right]\).}
    \label{fig:multi_params_analysis}
\end{figure}

As shown in FIG. \ref{fig:multi_params_analysis}(a), the AMPPI algorithm demonstrates efficient search strategies in three-dimensional parameter space \((T_c, \beta, \gamma)\) with the correlation length exponent fixed at \(\nu=1\). Multiple trajectories starting from different initial points all exhibit typical patterns similar to manual analysis by physicists: first quickly determining \(T_c\) along that direction, then synchronously optimizing critical exponents \(\beta\) and \(\gamma\) on an approximately constant \(T_c\) plane. This staged strategy avoids ineffective exponent optimization at inaccurate temperatures, significantly improving search efficiency and convergence speed.

Traditional finite-size scaling analysis typically requires large amounts of pre-collected data points, and parameter estimation processes often depend on manual experience. In contrast, the AMPPI algorithm realizes efficient parameter space exploration through real-time interaction with the physical environment, continuously deepening its understanding of the physical model. As shown in FIG. \ref{fig:multi_params_analysis}(b), as the iteration process progresses, the reward function quickly converges to its maximum value, while the three critical parameter estimates stabilize near theoretical expectations with continuously decreasing statistical errors. After 200 iterations, the fixed-\(\nu\) optimization provides high-precision critical parameter estimates of \(T_c=2.2690(3)\), \(\beta=0.1247(9)\), and \(\gamma=1.7490(43)\), with relative deviations less than 0.25\% from the exact solutions for the two-dimensional Ising model: \(T_c=2.269185...\), \(\beta=1/8=0.125\), and \(\gamma=7/4=1.75\).

FIG. \ref{fig:multi_params_analysis}(c) demonstrates the effectiveness of the auxiliary reward mechanism in resolving the correlation length exponent \(\nu\). The auxiliary reward \(R_{\mathrm{aux}}\) successfully enables determination of individual critical exponents rather than just their ratios, overcoming the non-uniqueness problem inherent in traditional finite-size scaling analysis. However, this capability comes with increased parameter variance, particularly affecting \(\gamma\) estimation precision, due to accumulated errors from the dynamic model's imperfect simulation of critical fluctuations.

We performed detailed comparisons between the AMPPI algorithm and the standard pyfssa algorithm (see Section \ref{subsec:detailed_multi_param_results}). The AMPPI algorithm demonstrated significant advantages in all tested system size combinations. Notably, after just 20 iterations, AMPPI's parameter estimation precision was already markedly superior to pyfssa's best results. With system sizes \(L=(32,64)\), AMPPI's results showed relative deviations less than 0.5\% from theoretical values, with significantly smaller statistical errors than pyfssa. Similar advantages were evident across all tested system size combinations.

The excellent performance of the AMPPI algorithm stems from two key mechanisms: first, it learns dynamic changes in physical observables through continuous interaction, autonomously constructing internal representations of critical behavior to achieve high-precision parameter estimation; second, the adaptive variance control mechanism creates "symmetry breaking" in optimal action generation (see Section \ref{subsec:explore_analyze}), enabling the algorithm to quickly identify target parameter directions from any initial point, greatly accelerating the search process.

The AMPPI algorithm demonstrates significant advantages over traditional methods: it eliminates dependence on initial parameter selection and enables simultaneous determination of multiple critical parameters through real-time environment interaction. However, FIG. \ref{fig:multi_params_analysis}(c) reveals the current limitation of the auxiliary reward approach: while it successfully resolves the non-uniqueness problem inherent in the global reward function \(R_{\mathrm{global}}\), enabling determination of individual critical exponents rather than just their ratios, it introduces increased parameter variance due to prediction errors from the dynamic model near the critical temperature.

\subsection{Transfer Learning} 

The Ising model exhibits similar physical mechanisms across different lattice structures, despite differences in critical parameters. This similarity motivated us to investigate whether knowledge gained from training on square lattice models could effectively transfer to triangular lattices. To assess this cross-lattice knowledge transfer capability, we designed two types of experiments: direct transfer and sample fine-tuning.

\subsubsection{Direct Transfer}

In the direct transfer experiment, we directly applied the AMPPI model trained on the square lattice Ising model to triangular lattice systems, maintaining the same initial parameter search range. FIG. \ref{fig:transfer_learning_analysis}(a) shows that the parameter estimates generated by the model (blue circles) are somewhat scattered, reflecting differences in thermodynamic behavior between the two lattice structures. Nevertheless, these estimates still form clusters around theoretical values, indicating that the model retains an understanding of basic Ising system characteristics.

\subsubsection{Sample Fine-Tuning Optimization}

\begin{figure}[htbp]
    \centering
    \includegraphics[width=\columnwidth]{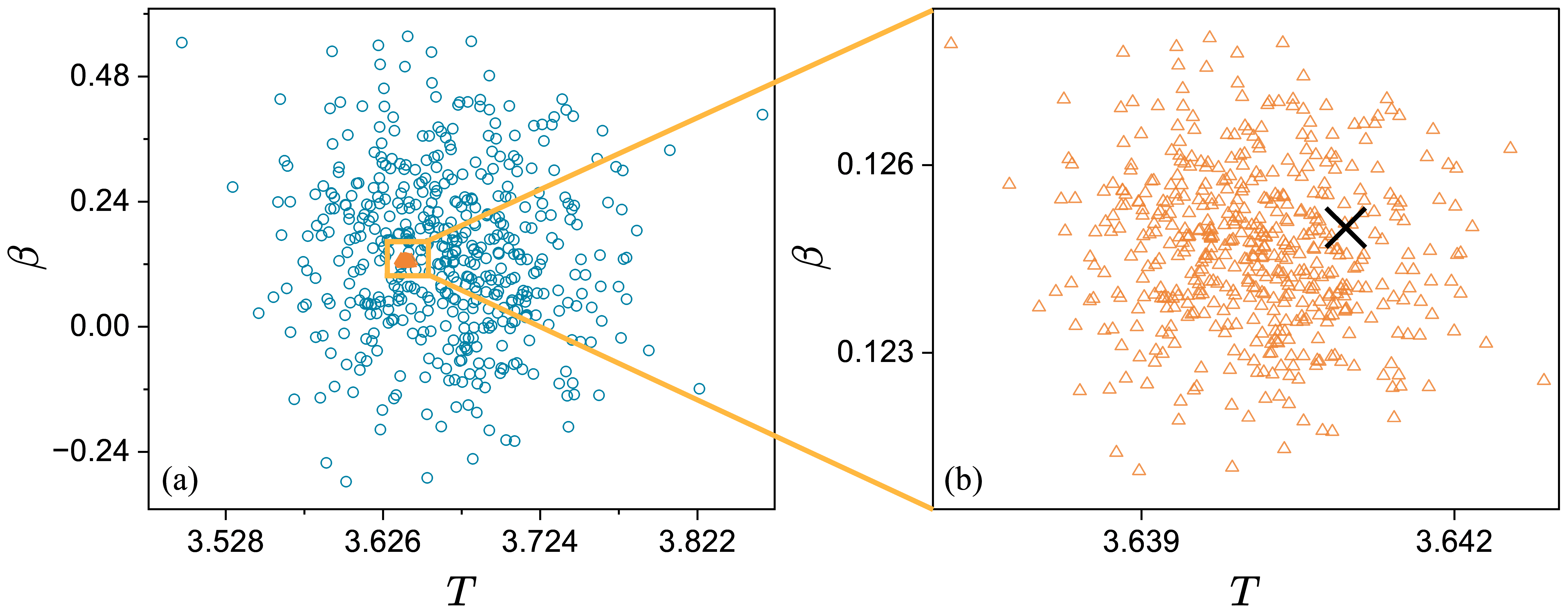}
    \captionsetup{justification=raggedright, singlelinecheck=false}
    \caption{Transfer learning results of the AMPPI algorithm in finding critical parameters ($T_c$, $\beta$) in the two-dimensional triangular lattice Ising model (\(L=(32, 64)\)). Blue circles in (a) represent the distribution of critical parameter values found after direct transfer, (b) shows a locally magnified area, where orange triangles represent the distribution of critical parameter values found after 20 iterations of fine-tuning based on physical environment sample data, and black crosses mark theoretical values ($T_c = \frac{4}{\ln(3)}$, $\beta = 0.125$).}
    \label{fig:transfer_learning_analysis}
\end{figure}

To improve model performance on triangular lattices, we fine-tuned the model using triangular lattice sample data. FIG. \ref{fig:transfer_learning_analysis}(b) demonstrates the significant effect after 20 iterations of fine-tuning: parameter estimates become markedly concentrated near theoretical values. After fine-tuning, the uncertainty of critical temperature $T_c$ decreased from $\pm0.0510$ to $\pm0.0009$, and the uncertainty of critical exponent $\beta$ from $\pm0.1602$ to $\pm0.0013$, improving precision by approximately 56 and 123 times, respectively. Detailed experimental setup and results are in Section \ref{subsec:detailed_transfer_learning_results}.

These results validate the AMPPI algorithm's transfer learning capability between different lattice structure Ising models. Through fine-tuning strategies, the model can quickly and accurately capture characteristics of new lattice structures, reducing parameter estimation uncertainties by two orders of magnitude.

This fine-tuning process offers a significant computational advantage. Training a new model from scratch necessitates an initial exploration phase (approximately 2,000 simulation steps in our setup) to build a global model. In contrast, our fine-tuning approach leverages the pre-trained model to bypass this expensive initialization entirely. By achieving high-precision convergence within just 20 iterations of the exploitation phase, the strategy demonstrates that transfer learning effectively eliminates the need for extensive random exploration, substantially reducing the total computational cost.

\section{Conclusion}
\label{conclusion}

This study presents a physics-inspired reinforcement learning method achieving high-precision autonomous discovery of the Ising model's critical parameters. By analogizing action variance to temperature changes, the algorithm exhibits phase transition-like exploration behavior, rapidly locating target parameters from any initial condition. Experimental results show that compared to traditional analysis methods, this approach achieves precision more than an order of magnitude higher while surpassing conventional reinforcement learning algorithms in efficiency. The composite reward structure combining global and auxiliary reward functions enables simultaneous determination of multiple critical parameters including the correlation length exponent \(\nu\), though this enhanced capability introduces increased parameter variance due to prediction errors from the dynamic model near the critical temperature, particularly affecting the precision of \(\gamma\) estimation. Additionally, the algorithm demonstrates excellent transfer learning capabilities that, combined with sample fine-tuning, effectively transfer knowledge to systems with different lattice structures. This research establishes a new paradigm for physical research, transitioning from manual analysis to AI-driven autonomous discovery. Based on the algorithm's demonstrated autonomous understanding capability and physics-inspired design, this method holds promise for application to more complex physical systems, particularly non-equilibrium systems and complex phase transitions that are difficult to address analytically. Future work will explore extending this method to quantum many-body systems, topological phase transitions, and critical dynamics.

\vskip 0.2in
\bibliography{ADIP}

\begin{thebibliography}{10}

\bibitem{botevCrossentropyMethodOptimization2013}
Zdravko~I. Botev, Dirk~P. Kroese, Reuven~Y. Rubinstein, and Pierre L'Ecuyer.
\newblock The cross-entropy method for optimization.
\newblock In {\em Handbook of {{Statistics}}}, volume~31, pages 35--59. Elsevier, 2013.

\bibitem{williamsModelPredictivePath2015}
Grady Williams, Andrew Aldrich, and Evangelos Theodorou.
\newblock Model {{Predictive Path Integral Control}} using {{Covariance Variable Importance Sampling}}, October 2015.

\bibitem{williamsAggressiveDrivingModel2016}
Grady Williams, Paul Drews, Brian Goldfain, James~M. Rehg, and Evangelos~A. Theodorou.
\newblock Aggressive driving with model predictive path integral control.
\newblock In {\em 2016 {{IEEE International Conference}} on {{Robotics}} and {{Automation}} ({{ICRA}})}, pages 1433--1440. IEEE, May 2016.

\bibitem{williamsInformationTheoreticMPC2017}
Grady Williams, Nolan Wagener, Brian Goldfain, Paul Drews, James~M. Rehg, Byron Boots, and Evangelos~A. Theodorou.
\newblock Information theoretic {{MPC}} for model-based reinforcement learning.
\newblock In {\em 2017 {{IEEE International Conference}} on {{Robotics}} and {{Automation}} ({{ICRA}})}, pages 1714--1721. IEEE, May 2017.

\bibitem{nagabandiDeepDynamicsModels2019}
Anusha Nagabandi, Kurt Konoglie, Sergey Levine, and Vikash Kumar.
\newblock Deep {{Dynamics Models}} for {{Learning Dexterous Manipulation}}, September 2019.

\bibitem{kappenLinearTheoryControl2005}
Hilbert~J. Kappen.
\newblock Linear {{Theory}} for {{Control}} of {{Nonlinear Stochastic Systems}}.
\newblock {\em Physical Review Letters}, 95(20):200201, November 2005.

\bibitem{kappenPathIntegralsSymmetry2005}
H~J Kappen.
\newblock Path integrals and symmetry breaking for optimal control theory.
\newblock {\em Journal of Statistical Mechanics: Theory and Experiment}, 2005(11):P11011--P11011, November 2005.

\bibitem{wolffCollectiveMonteCarlo1989}
Ulli Wolff.
\newblock Collective {{Monte Carlo Updating}} for {{Spin Systems}}.
\newblock 62(4):361--364.

\bibitem{sorgePyfssa0762015}
Andreas Sorge.
\newblock Pyfssa 0.7.6.
\newblock Zenodo, December 2015.

\bibitem{nelderSimplexMethodFunction1965}
J.~A. Nelder and R.~Mead.
\newblock A {{Simplex Method}} for {{Function Minimization}}.
\newblock {\em The Computer Journal}, 7(4):308--313, January 1965.

\bibitem{houdayerLowtemperatureBehaviorTwodimensional2004}
J{\'e}r{\^o}me Houdayer and Alexander Hartmann.
\newblock Low-temperature behavior of two-dimensional {{Gaussian Ising}} spin glasses.
\newblock {\em Physical Review B}, 70(1):014418, July 2004.

\end{thebibliography}


\begin{thebibliography}{10}

\bibitem{cardyFinitesizeScaling1988}
John~L. Cardy.
\newblock {\em Finite-Size Scaling}.
\newblock Number~2 in Current Physics. {North-Holland Sole distributors for the USA and Canada, Elsevier Science Pub. Co}, 1988.

\bibitem{binderFiniteSizeScaling1981}
K.~Binder.
\newblock Finite size scaling analysis of ising model block distribution functions.
\newblock {\em Zeitschrift f{\"u}r Physik B Condensed Matter}, 43(2):119--140, June 1981.

\bibitem{selkeCriticalBinderCumulant2006}
W.~Selke.
\newblock Critical {{Binder}} cumulant of two-dimensional {{Ising}} models.
\newblock {\em The European Physical Journal B - Condensed Matter and Complex Systems}, 51(2):223--228, May 2006.

\bibitem{bhattacharjeeMeasureDataCollapse2001}
Somendra~M Bhattacharjee and Flavio Seno.
\newblock A measure of data collapse for scaling.
\newblock {\em Journal of Physics A: Mathematical and General}, 34(33):6375--6380, August 2001.

\bibitem{wenzelPercolationVortices3D2008}
Sandro Wenzel, Elmar Bittner, Wolfhard Janke, and Adriaan~M.J. Schakel.
\newblock Percolation of vortices in the {{3D Abelian}} lattice {{Higgs}} model.
\newblock {\em Nuclear Physics B}, 793(1-2):344--361, April 2008.

\bibitem{melchertAutoScalepyProgramAutomatic2009}
O.~Melchert.
\newblock {{autoScale}}.py - a program for automatic finite-size scaling analyses: {{A}} user's guide, 2009.

\bibitem{sorgePyfssa0762015}
Andreas Sorge.
\newblock Pyfssa 0.7.6.
\newblock Zenodo, December 2015.

\bibitem{klockeTopologicalOrderEntanglement2022}
Kai Klocke and Michael Buchhold.
\newblock Topological order and entanglement dynamics in the measurement-only {{XZZX}} quantum code.
\newblock {\em Physical Review B}, 106(10):104307, September 2022.

\bibitem{carrasquillaMachineLearningPhases2017}
Juan Carrasquilla and Roger~G. Melko.
\newblock Machine learning phases of matter.
\newblock {\em Nature Physics}, 13(5):431--434, May 2017.

\bibitem{chngMachineLearningPhases2017}
Kelvin Ch'ng, Juan Carrasquilla, Roger~G. Melko, and Ehsan Khatami.
\newblock Machine learning phases of strongly correlated fermions.
\newblock {\em Physical Review X}, 7(3):031038, August 2017.

\bibitem{vannieuwenburgLearningPhaseTransitions2017}
Evert~P.~L. Van~Nieuwenburg, Ye-Hua Liu, and Sebastian~D. Huber.
\newblock Learning phase transitions by confusion.
\newblock {\em Nature Physics}, 13(5):435--439, May 2017.

\bibitem{suchslandParameterDiagnosticsPhases2018}
Philippe Suchsland and Stefan Wessel.
\newblock Parameter diagnostics of phases and phase transition learning by neural networks.
\newblock {\em Physical Review B}, 97(17):174435, May 2018.

\bibitem{zhangFewshotMachineLearning2019}
Rui Zhang, Bin Wei, Dong Zhang, Jia-Ji Zhu, and Kai Chang.
\newblock Few-shot machine learning in the three-dimensional {{Ising}} model.
\newblock {\em Physical Review B}, 99(9):094427, March 2019.

\bibitem{zhangInterpretableMachineLearning2019}
Wei Zhang, Lei Wang, and Ziqiang Wang.
\newblock Interpretable machine learning study of the many-body localization transition in disordered quantum {{Ising}} spin chains.
\newblock {\em Physical Review B}, 99(5):054208, February 2019.

\bibitem{liExtractingCriticalExponents2019}
Zhenyu Li, Mingxing Luo, and Xin Wan.
\newblock Extracting critical exponents by finite-size scaling with convolutional neural networks.
\newblock {\em Physical Review B}, 99(7):075418, February 2019.

\bibitem{theveniautNeuralNetworkSetups2019}
Hugo Th{\'e}veniaut and Fabien Alet.
\newblock Neural network setups for a precise detection of the many-body localization transition: {{Finite-size}} scaling and limitations.
\newblock {\em Physical Review B}, 100(22):224202, December 2019.

\bibitem{raoMachineLearningManybody2020}
Wen-Jia Rao.
\newblock Machine learning for many-body localization transition*.
\newblock {\em Chinese Physics Letters}, 37(8):080501, August 2020.

\bibitem{funaiThermodynamicsFeatureExtraction2020}
Shotaro~Shiba Funai and Dimitrios Giataganas.
\newblock Thermodynamics and feature extraction by machine learning.
\newblock {\em Physical Review Research}, 2(3):033415, September 2020.

\bibitem{wangUnsupervisedLearningTopological2021}
Jielin Wang, Wanzhou Zhang, Tian Hua, and Tzu-Chieh Wei.
\newblock Unsupervised learning of topological phase transitions using the calinski-harabaz index.
\newblock {\em Physical Review Research}, 3(1):013074, January 2021.

\bibitem{rodriguesdeassiseliasGlobalExplorationPhase2022}
Danilo Rodrigues De Assis~Elias, Enzo Granato, and Maurice De~Koning.
\newblock Global exploration of phase behavior in frustrated ising models using unsupervised learning techniques.
\newblock {\em Physica A: Statistical Mechanics and its Applications}, 589:126653, March 2022.

\bibitem{yueIncrementalLearningPhase2022}
Zhenyi Yue, Yuqi Wang, and Pin Lyu.
\newblock Incremental learning of phase transition in ising model: {{Preprocessing}}, finite-size scaling and critical exponents.
\newblock {\em Physica A: Statistical Mechanics and its Applications}, 600:127538, August 2022.

\bibitem{yonedaNeuralNetworkApproach2023}
Ryosuke Yoneda and Kenji Harada.
\newblock Neural network approach to scaling analysis of critical phenomena.
\newblock {\em Physical Review E}, 107(4):044128, April 2023.

\bibitem{wolffCollectiveMonteCarlo1989}
Ulli Wolff.
\newblock Collective {{Monte Carlo Updating}} for {{Spin Systems}}.
\newblock 62(4):361--364.

\bibitem{botevCrossentropyMethodOptimization2013}
Zdravko~I. Botev, Dirk~P. Kroese, Reuven~Y. Rubinstein, and Pierre L'Ecuyer.
\newblock The cross-entropy method for optimization.
\newblock In {\em Handbook of {{Statistics}}}, volume~31, pages 35--59. Elsevier, 2013.

\bibitem{williamsModelPredictivePath2015}
Grady Williams, Andrew Aldrich, and Evangelos Theodorou.
\newblock Model {{Predictive Path Integral Control}} using {{Covariance Variable Importance Sampling}}, October 2015.

\bibitem{williamsAggressiveDrivingModel2016}
Grady Williams, Paul Drews, Brian Goldfain, James~M. Rehg, and Evangelos~A. Theodorou.
\newblock Aggressive driving with model predictive path integral control.
\newblock In {\em 2016 {{IEEE International Conference}} on {{Robotics}} and {{Automation}} ({{ICRA}})}, pages 1433--1440. IEEE, May 2016.

\bibitem{williamsInformationTheoreticMPC2017}
Grady Williams, Nolan Wagener, Brian Goldfain, Paul Drews, James~M. Rehg, Byron Boots, and Evangelos~A. Theodorou.
\newblock Information theoretic {{MPC}} for model-based reinforcement learning.
\newblock In {\em 2017 {{IEEE International Conference}} on {{Robotics}} and {{Automation}} ({{ICRA}})}, pages 1714--1721. IEEE, May 2017.

\bibitem{nagabandiDeepDynamicsModels2019}
Anusha Nagabandi, Kurt Konoglie, Sergey Levine, and Vikash Kumar.
\newblock Deep {{Dynamics Models}} for {{Learning Dexterous Manipulation}}, September 2019.

\bibitem{kappenLinearTheoryControl2005}
Hilbert~J. Kappen.
\newblock Linear {{Theory}} for {{Control}} of {{Nonlinear Stochastic Systems}}.
\newblock {\em Physical Review Letters}, 95(20):200201, November 2005.

\bibitem{kappenPathIntegralsSymmetry2005}
H~J Kappen.
\newblock Path integrals and symmetry breaking for optimal control theory.
\newblock {\em Journal of Statistical Mechanics: Theory and Experiment}, 2005(11):P11011--P11011, November 2005.

\bibitem{nelderSimplexMethodFunction1965}
J.~A. Nelder and R.~Mead.
\newblock A {{Simplex Method}} for {{Function Minimization}}.
\newblock {\em The Computer Journal}, 7(4):308--313, January 1965.

\bibitem{houdayerLowtemperatureBehaviorTwodimensional2004}
J{\'e}r{\^o}me Houdayer and Alexander Hartmann.
\newblock Low-temperature behavior of two-dimensional {{Gaussian Ising}} spin glasses.
\newblock {\em Physical Review B}, 70(1):014418, July 2004.

\end{thebibliography}

\appendix
% ****** Start of file Supplemental_Material_JSTAT_en.tex ******

% \documentclass[twoside,11pt]{article}

% \usepackage{blindtext}

% Any additional packages needed should be included after jmlr2e.
% Note that jmlr2e.sty includes epsfig, amssymb, natbib and graphicx,
% and defines many common macros, such as 'proof' and 'example'.
%
% It also sets the bibliographystyle to plainnat; for more information on
% natbib citation styles, see the natbib documentation, a copy of which
% is archived at http://www.jmlr.org/format/natbib.pdf

% Available options for package jmlr2e are:
% 
%   - abbrvbib : use abbrvnat for the bibliography style
%   - nohyperref : do not load the hyperref package
%   - preprint : remove JMLR specific information from the template,
%         useful for example for posting to preprint servers.
% 
% Example of using the package with custom options:
% 

\renewcommand{\theequation}{S.\arabic{equation}}
\renewcommand{\thetable}{S\arabic{table}}
\renewcommand{\thefigure}{S\arabic{figure}}
\renewcommand{\thealgorithm}{S\arabic{algorithm}}

\newcommand{\Input}{\item[\textbf{Input:}]}
\newcommand{\Output}{\item[\textbf{Output:}]}

% \title{Detailed Performance of Transfer Learning in Ising Model} % Add your title here

% \maketitle
% \tableofcontents
% \clearpage

% Environment
\section{Ising Model Environment}
\label{sec:ising_env}

This section establishes the physical environment required for the reinforcement learning framework. The Hamiltonian of the two-dimensional Ising model system can be expressed as:
\begin{equation}
    H = -J \sum_{\langle i,j \rangle} \sigma_i \sigma_j,
    \label{eq:supp_hamiltonian}
\end{equation}

In equation \ref{eq:supp_hamiltonian}, \( J > 0 \) represents the ferromagnetic coupling strength, \( \sigma_i \in \{+1, -1\} \) denotes the spin state at lattice site \(i\), and the summation is limited to nearest-neighbor site pairs \(\langle i,j \rangle\). To characterize the critical behavior of the system, we measure the following thermodynamic quantities: magnetization \( M = \frac{1}{N} \left\langle \left| \sum_i \sigma_i \right| \right\rangle \), energy \( E = \frac{1}{N} \langle H \rangle \), specific heat \( C_v = \frac{1}{NT^2} \left( \langle H^2 \rangle - \langle H \rangle^2 \right) \), magnetic susceptibility \( \chi = \frac{1}{NT} \left( \langle M^2 \rangle - \langle M \rangle^2 \right) \), and Binder cumulant \( U_L = 1 - \frac{\langle M^4 \rangle}{3 \langle M^2 \rangle^2} \). All these physical quantities constitute the observation set \( O_L = \{ M_L, E_L, C_{v,L}, \chi_L, U_L \} \), where the subscript \(L\) indicates the linear size of the system.

Near the critical point \( T_c \), the observables follow standard finite-size scaling theory. The universal scaling relation can be expressed as:

\begin{equation}
    O_L = L^{\kappa/\nu} \tilde{O}(tL^{1/\nu}),
    \label{eq:finite_size_scaling}
\end{equation}

In equation \ref{eq:finite_size_scaling}, the reduced temperature parameter \( t = (T-T_c)/T_c \) measures the deviation of the system from the critical point, and \( \tilde{O} \) represents the universal scaling function. The critical scaling exponent \( \kappa \) varies with different observables: for magnetization \(M\), \(\kappa=-\beta\); for magnetic susceptibility \(\chi\), \(\kappa=\gamma\); and for the Binder cumulant \(U_L\), \(\kappa=0\). The parameter \( \nu \) characterizes the scaling behavior of the correlation length.

Based on the observables and their scaling relations, the state space of the reinforcement learning environment is defined as the set of physical observations and their scaled forms:

\begin{equation}
    \mathcal{S} = \left\{ \left( O_L, \tilde{O}_L \right) \mid L \in \mathbb{Z}^+ \right\},
    \label{eq:supp_state_space}
\end{equation}

The scaled form of the observables in the state space is defined as \( \tilde{O}_L = O_L \cdot L^{-\kappa/\nu} \). The corresponding action space allows the agent to adjust key physical parameters:

\begin{equation}
\mathcal{A} = \left\{ \Delta\varphi = (\Delta T, \Delta\beta, \Delta\gamma, \Delta\nu) \mid \Delta T \in \mathbb{R},\ \Delta\beta \in \mathbb{R},\ \Delta\gamma \in \mathbb{R}, \Delta\nu \in \mathbb{R} \right\}.
\label{eq:action_space}
\end{equation}

This continuous action space enables the agent to fine-tune the temperature \(T\) and critical exponents \(\beta\), \(\gamma\) and \(\nu\) in each iteration step.

The agent seeks critical parameters $\varphi = (T, \beta, \gamma, \nu)$ that maximize data collapse quality for physical observables across different system sizes. We define a general reward formulation as:

\begin{equation}
    \mathcal{R} = e^{-d({\mathcal{N}(\tilde{\mathcal{O}})})}
     \label{eq:supp_general_reward}
\end{equation}

Here, \(d({\mathcal{N}(\tilde{\mathcal{O}})}) = \sqrt{\sum_{i=1}^{n-1} (\mathcal{N}(\tilde{\mathcal{O}})_{L{i+1}} - \mathcal{N}(\tilde{\mathcal{O}})_{L_i})^2}\) quantifies the distance between normalized scaled observables of different system sizes. We apply Min-Max normalization: \(\mathcal{N}(\mathcal{O}_L) = \frac{\mathcal{O}_L - \mathcal{O}_{\min}}{\mathcal{O}_{\max} - \mathcal{O}_{\min}}\), where \(\mathcal{O}_{\min}\) and \(\mathcal{O}_{\max}\) are the minimum and maximum values across different system sizes. The scaled observables \(\tilde{\mathcal{O}}_L\) represent physical quantities transformed according to finite-size scaling theory (e.g., \(\tilde{M}_L = M_L L^{\beta/\nu}\) for magnetization). When critical parameters are correct, the scaled observables collapse onto universal curves, the distance \(d\) approaches zero, and the reward reaches its maximum value of 1.

The correlation length exponent \(\nu\) requires special treatment due to inherent non-uniqueness issues, as detailed in the following section.

% Methodological Foundations
\section{Computational Methods}
\label{sec:methods}
\subsection{Markov Decision Process (MDP)}
\label{subsec:mdp_formulation}
% Formal definition of state/action spaces
Markov Decision Process (MDP) provides the foundational framework for reinforcement learning. Formally, an MDP is defined as a quintuple:
\begin{equation}
\mathcal{M} = (\mathcal{S}, \mathcal{A}, \mathcal{P}, \mathcal{R}, \gamma_d),
\label{eq:mdp_tuple}
\end{equation}

Here, \(\mathcal{S} \) represents the set of all possible system states, containing the observables defined by equation \ref{eq:supp_state_space}; \(\mathcal{A} \) is the set of actions, given by equation \ref{eq:action_space}; the transition function \(\mathcal{P}(s_{t+1}|s_t,a_t) \) characterizes the system dynamics, describing the probability of transitioning to state \(s_{t+1} \) given the current state-action pair \((s_t,a_t)\); the reward function \(\mathcal{R}(s_t,a_t) \) quantifies the immediate return for each state-action combination; and the discount factor \(\gamma_d \in [0,1] \) balances the importance of immediate versus long-term rewards.

At each time step \(t\), the agent observes the current state \(s_t \in \mathcal{S}\), selects an action \(a_t \in \mathcal{A}\), and the system generates the next state \(s_{t+1} \) according to the transition probability \(\mathcal{P} \) and provides an immediate reward \(r_t = \mathcal{R}(s_t, a_t)\). This process optimizes the policy through accumulated rewards over time steps.

The agent selects actions through a policy \(\pi\), which is a mapping from the state space to the action space:
\begin{equation}
\pi: \mathcal{S} \to \mathcal{A},
\label{eq:policy_mapping}
\end{equation}

The policy \(\pi \) can be deterministic or stochastic. Within the MDP framework, the agent's objective is to maximize its cumulative return by optimizing the policy \(\pi\). Given a policy \(\pi\), we define the state-action value function \(Q^\pi(s, a)\), which represents the expected return when taking action \(a \) in state \(s \) and following policy \(\pi \) thereafter:
\begin{equation}
Q^\pi(s, a) = \mathbb{E}\left[\sum_{h=0}^{H} \gamma_d^h r_{t+h} \mid s_t = s, a_t = a \right],
\label{eq:q_value_function}
\end{equation}
where \(H \) is the trajectory length and \(\gamma_d \) is the discount factor.

The optimal policy \(\pi^* \) can be obtained by maximizing \(Q^\pi(s, a)\):
\begin{equation}
\pi^* = \arg \max_{\pi} Q^\pi(s, a) = \arg \max_{\pi} \mathbb{E}\left[\sum_{h=0}^{H} \gamma_d^h r_{t+h} \mid s_t = s, a_t = a \right].
\label{eq:optimal_policy}
\end{equation}
Through the optimal policy, the agent selects actions that maximize the cumulative return, generally defined as the discounted return \(R=\sum_{h=0}^{H} \gamma_d^h r_{t+h}\).

MDP provides a fundamental framework for reinforcement learning, where the definitions of state space and action space, along with the mechanisms of state transition and reward feedback, ensure that the agent can optimize its behavior through environmental exploration.

\subsection{Model-Based Reinforcement Learning (MBRL)}

\label{subsec:mbrl_arch}
Model-Based Reinforcement Learning (MBRL) is a reinforcement learning approach where the agent constructs a dynamic model of the environment during the learning process and utilizes this model for planning and decision-making. Unlike value function-based methods, MBRL learns an environment model to predict system state transitions, which serves as the basis for action selection and policy optimization.

In MBRL, the dynamic transition model of the environment can be represented as \(p_{\theta}(s_{t+1} | s_t, a_t)\), which describes the probability distribution of the system transitioning to the next state \(s_{t+1} \) given the current state \(s_t \) and action \(a_t\). In this study, we use deep neural networks to approximate this transition function, with network parameters \(\theta \) continuously updated through the training process.

\subsubsection{Model Learning}
\label{subsubsec:model_learning}

Dynamic system modeling is a core element of MBRL algorithms. By fitting the state transition probability distribution with neural networks, the complex dynamics of physical systems can be represented as a conditional probability model:

\begin{equation}
p_{\theta}(s_{t+1} | s_t, a_t) = \mathcal{N}(f_\theta(s_t, a_t), \Sigma_\mathrm{model}),
\label{eq:state_transition_model}
\end{equation}

Equation \ref{eq:state_transition_model} models the state transition process using a multivariate Gaussian distribution, where the mean vector \(f_\theta(s_t, a_t) \) is provided by a parameterized neural network, and the covariance matrix \(\Sigma_\mathrm{model} \) characterizes the statistical uncertainty of the prediction. This probabilistic modeling approach can capture both the deterministic components and random fluctuations of system dynamics.

The time series characteristics of physical systems require models capable of capturing long-term dependencies, and recurrent neural network (RNN) architectures excel in this aspect. The function \(f_{\theta}(s, a) \) implemented as an RNN processes temporal dependencies through a recursive structure, with hidden state updates following:

\begin{equation}
h_t = \tanh(W_{ih}x_t + b_{ih} + W_{hh}h_{t-1} + b_{hh}),
\label{eq:rnn_hidden_state}
\end{equation}

In the RNN computational graph, the input vector \(x_t \) includes the feature representation of the current state-action pair \((s_t,a_t)\), and \(h_t \) represents the hidden state vector at time \(t\). The matrix \(W_{ih} \) connects the input layer with the hidden layer, while \(W_{hh} \) implements information transfer across the time dimension. The vectors \(b_{ih} \) and \(b_{hh} \) are the corresponding bias terms. The hyperbolic tangent activation function \(\tanh(\cdot) \) introduces non-linearity, enabling the network to capture complex patterns in system dynamics. The parameter set \(\theta = \{W_{ih}, W_{hh}, b_{ih}, b_{hh}\} \) is optimized through gradient descent methods.

To enhance prediction accuracy and robustness, we employ model ensemble techniques. Based on statistical learning theory, combining multiple independently trained models can effectively reduce prediction variance and improve generalization capability. The ensemble prediction value is calculated as:

\begin{equation}
f\left(s_t, a_t\right)=\frac{1}{N} \sum_{n=1}^N f_{\theta_n}\left(s_t, a_t\right)
\label{eq:ensemble_prediction}
\end{equation}

Equation \ref{eq:ensemble_prediction} implements the arithmetic mean of predictions from \(N \) independently trained models. Each prediction function \(f_{\theta_n} \) has a unique parameter set \(\theta_n \) obtained through random initialization and different training data subsets. This ensemble strategy significantly enhances prediction reliability in regions of higher uncertainty, such as physical phase transition regions.

In practical implementation, we adopted a multi-stage ensemble model training strategy to improve model prediction accuracy and generalization capability. Specifically, ensemble models are divided into two groups: one group consists of global models (total \(N_g\)), primarily trained on trajectory data sampled from random policies, enabling them to describe dynamic changes across a broader state space; the other group consists of local models (total \(N_l\)), specifically trained on trajectory data generated by planning algorithms, enabling them to provide more precise descriptions of state space changes in small regions near target parameters.

This hierarchical training strategy offers several significant advantages: first, global models, by learning a wide range of state transitions, provide the agent with a reliable understanding of the overall environment dynamics, supporting exploration in the initial phase; second, local models, by focusing on state changes near critical points, provide higher precision local predictions, thereby improving algorithm performance in key regions; finally, this strategy, by combining macro-level understanding and micro-level precision, enables the algorithm to work efficiently and robustly under any initial conditions.

To further optimize the ensemble strategy, we designed a hierarchical weighted ensemble architecture that combines the advantages of global and locally specialized models:

\begin{equation}
f(s_t, a_t) = \alpha_M \cdot \frac{1}{N_g} \sum_{n=1}^{N_g} f_{\theta_n^g}(s_t, a_t) + (1-\alpha_M) \cdot \frac{1}{N_l} \sum_{m=1}^{N_l} f_{\theta_m^l}(s_t, a_t)
\label{eq:ensemble_prediction_weighted}
\end{equation}

Here, the first term represents the average prediction of \(N_g \) global models, which excel at capturing general dynamic features across a broad state space; the second term represents the contribution of \(N_l \) local models, focusing on fine-grained behavior near critical regions. The dynamic weight coefficient \(\alpha_M \in [0,1] \) adaptively adjusts the relative importance of the two types of models based on the system state, thereby achieving a balance between exploration and precise estimation.

The training objective for the dynamic model is to minimize the difference between predicted state changes and actual state changes. For a given dataset \(\mathcal{D}=\left\{s_t, a_t, s_{t+1}\right\}\), we use mean squared error as the loss metric:
\begin{equation}
    \mathcal{L}(\theta)=\frac{1}{|\mathcal{D}|} \sum_{\left(s_t, a_t, s_{t+1}\right) \in \mathcal{D}} \frac{1}{2}\left\|\left(s_{t+1}-s_t\right)-f_\theta\left(s_t, a_t\right)\right\|_2^2.
    \label{eq:loss_function_unnormalized}
\end{equation}

The Euclidean distance squared between the true state change \(\Delta s_{\mathrm{true}} = s_{t+1}-s_t \) and the predicted state change \(\Delta s_{\mathrm{pred}} = f_\theta\left(s_t, a_t\right) \) quantifies the model error. This differential learning strategy is more stable compared to directly predicting absolute state values and can better capture small changes in critical regions.

Considering that physical observables have significant differences in order of magnitude (e.g., magnetic susceptibility \(\chi \) and Binder cumulant \(U_L\)), we introduce a data normalization strategy to improve training dynamics. The modified loss function is expressed as:

\begin{equation}
    \mathcal{L}(\theta)=\frac{1}{|\mathcal{D}|} \sum_{\left(s_t, a_t, s_{t+1}\right) \in \mathcal{D}} \frac{1}{2}\left\| \mathrm{ Normalize }\left(s_{t+1}-s_t\right)-f_\theta\left(s_t, a_t\right) \right\|_2^2.
    \label{eq:loss_function_normalized}
\end{equation}

The normalization transformation \(\mathrm{Normalize}(x) = \frac{x - \mu_x}{\sigma_x} \) projects the state change vector to a distribution space with zero mean and unit standard deviation, where \(\mu_x \) and \(\sigma_x \) are the mean and standard deviation vectors of state changes in the training set, respectively. This preprocessing technique significantly enhances training stability, especially when handling rapid changes in physical quantities near critical points. During the prediction phase, we apply the inverse transformation \(s_{t+1} = s_t + \sigma_x \cdot f_{\theta}(s_t, a_t) + \mu_x \) to restore the original scale.

To improve training efficiency and prevent overfitting, we implemented a dynamic early stopping mechanism based on validation set performance, adaptively determining the optimal number of training epochs. The decision criterion is formalized as:

\begin{equation}
    \mathrm{Stop if } \min_{i \in \{t-p, \ldots, t\}} \mathcal{L}_{\mathrm{val}}^{(i)} > \min_{j \in \{1, \ldots, t-p-1\}} \mathcal{L}_{\mathrm{val}}^{(j)} - \epsilon
    \label{eq:early_stopping}
\end{equation}

Equation \ref{eq:early_stopping} implements a sliding window comparison mechanism: training terminates when the best validation loss value in the most recent \(p \) epochs exceeds the historical best loss value minus a tolerance threshold \(\epsilon\). The patience parameter \(p \) is set to \(40\), and the tolerance threshold \(\epsilon=10^{-4}\); this configuration balances model generalization capability and training efficiency.

These training strategies significantly improve the efficiency and stability of model training, enabling the model to achieve good performance in a shorter training time while avoiding overfitting issues.

\subsubsection{Model Predictive Control (MPC)}
\label{subsubsec:mpc}
% MPC implementation specifics
Model Predictive Control (MPC) is a control method that utilizes an environment model to predict future states and selects the optimal action sequence based on the predicted results. In the MPC framework, the agent first uses a dynamic model to predict the state evolution trajectory for a future period starting from the current state, then optimizes the selection of control actions in the prediction horizon based on a predefined reward function.

Specifically, MPC determines the optimal control sequence \(A^* \) by solving the following optimization problem:
\begin{equation}
A^* = \arg \max_{\{A^{(0)}, \ldots, A^{(N-1)}\}} \sum_{h=0}^{H-1} r(\hat{s}_{t+h}, a_{t+h}) \quad s.t. \quad \hat{s}_{t+1} = \hat{s}_t + f_\theta(\hat{s}_t, a_t),
\label{eq:mpc_optimization}
\end{equation}

Here, \(A^{(k)} = \{a_t^{(k)}, a_{t+1}^{(k)}, \ldots, a_{t+H-1}^{(k)}\} \) represents the \(k\)-th of \(N \) candidate action sequences, with each sequence containing \(H \) consecutive actions across time steps. The constraint enforces system evolution laws through the learned dynamic model \(f_\theta\), ensuring that the predicted state sequence \(\{\hat{s}_t, \hat{s}_{t+1}, \ldots, \hat{s}_{t+H}\} \) follows the physical system dynamics. The optimal sequence \(A^* \) is obtained by maximizing the cumulative reward value \(r(\hat{s}_{t+h}, a_{t+h})\).

\subsection{Cross Entropy Method (CEM)}
\label{subsec:cem_impl}
% CEM algorithm parameters
The Cross Entropy Method (CEM) is an iterative optimization algorithm that seeks optimal solutions by updating sampling distribution parameters\cite{botevCrossentropyMethodOptimization2013}. Its core idea is to continuously adjust probability distribution parameters during iterations to increase the sampling probability of high-quality solutions. Each iteration includes three key steps: sampling, evaluation, and distribution update.

In the \(m\)-th iteration, CEM generates \(N \) sets of candidate action sequences \(A = \{A_1, \ldots, A_N\} \) from a multivariate Gaussian distribution \(\mathcal{N}(\mu^m, \Sigma^m)\). Each action sequence \(A_i \) contains actions \(\{a_0^i, \ldots, a_{H-1}^i\} \) for \(H \) time steps, where at each time step \(t\), action \(a_t^i \) is independently sampled from the Gaussian distribution \(\mathcal{N}(\mu_t^m, \Sigma_t^m)\), in the specific form:

\begin{equation}
    \label{eq:cem_sampling}
    \begin{array}{ll}
      A_i = \{a_0^i, \ldots, a_{H-1}^i\}, \quad \mbox{where} & \quad a_t^i \sim \mathcal{N}(\mu_t^m, \Sigma_t^m), \\
    & \forall i \in \{1, \ldots, N\}, t \in \{0, \ldots, H-1\}
    \end{array}
\end{equation}

The distribution parameters \(\mu^m = \{\mu_0^m, \ldots, \mu_{H-1}^m\} \) and \(\Sigma^m = \{\Sigma_0^m, \ldots, \Sigma_{H-1}^m\} \) represent the means and covariances of actions at each time step in the \(m\)-th iteration, respectively.

After obtaining candidate action sequences, the algorithm enters the evaluation step, calculating performance metric \(J_i \) for each candidate action sequence \(A_i\). This metric is obtained by inputting the action sequence into the dynamic model \(f_\theta \) and accumulating reward values on the predicted trajectory:
\begin{equation}
    J_i = \sum_{t=0}^{H-1} r(\hat{s}_{i,t}, a_t^i), \quad \mathrm{where} \quad \hat{s}_{i,t+1} = \hat{s}_{i,t} + f_\theta(\hat{s}_{i,t}, a_t^i)
\end{equation}
    
In this formula, \(r(\hat{s}_{i,t}, a_t^i) \) represents the immediate reward obtained by executing action \(a_t^i \) in the predicted state \(\hat{s}_{i,t}\), and \(\hat{s}_{i,t+1} \) is the next state predicted by the dynamic model \(f_\theta\).
    
After completing the evaluation, the CEM algorithm selects the \(E \) most outstanding action sequences to form the elite set \(A_{elites}\). The elite set is selected as follows:
\begin{equation}
   A_{elites} = \mathrm{sort}(A)[-E:],
\end{equation}

Here, \(\mathrm{sort}(A) \) represents sorting all action sequences \(A \) in ascending order according to their performance metrics \(J_i\), and selecting the top \(E \) sequences (those with the highest performance) to form the elite set \(A_{elites}\).

Finally, the mean \(\mu^{m+1} \) and covariance \(\Sigma^{m+1} \) for the next iteration are updated using the elite set \(A_{elites}\). To ensure the stability of parameter updates, an exponential moving average method is typically employed:
\begin{equation}
   \mu_t^{m+1} = \alpha_{CEM} \cdot \mathrm{mean}(A_{elites, t}) + (1 - \alpha_{CEM})\mu_t^m, \quad \forall t \in \{0, \ldots, H-1\}
\end{equation}
\begin{equation}
   \Sigma_t^{m+1} = \alpha_{CEM} \cdot \mathrm{var}(A_{elites, t}) + (1 - \alpha_{CEM})\Sigma_t^m, \quad \forall t \in \{0, \ldots, H-1\}
\end{equation}

Where \(\mathrm{mean}(A_{elites, t}) \) and \(\mathrm{var}(A_{elites, t}) \) calculate the mean and variance of actions in the elite set \(A_{elites} \) at time step \(t\), respectively. The parameter \(\alpha_{CEM} \in [0, 1] \) is a smoothing coefficient that controls the magnitude of distribution parameter updates; a larger value makes updates more inclined toward the statistical properties of current elite samples, while a smaller value retains more historical information.

After \(M \) rounds of iterations, the final mean vector \(\mu = \mu^{M+1} \) is considered the optimal action sequence for the current optimization problem and can be directly applied to the actual control of the physical system.

\subsection{Adaptive Model Predictive Path Integral (AMPPI) Algorithm}
\label{subsec:amppi_detail}
% AMPPI core algorithm architecture
Adaptive Model Predictive Path Integral Control (AMPPI) algorithm combines the Path Integral Control (PIC) method with the Reward-Error Adaptive Variance Control (REAVC) strategy proposed in this paper based on physical intuition. It aims to overcome the limitation of traditional path integral control methods that easily fall into local optima, and to enhance the efficiency and precision of parameter exploration in complex physical environments. The following sections detail the composition and core mechanisms of the AMPPI algorithm.

\subsubsection{Path integral control (PIC)}
\label{subsubsec:pic}
Path Integral Control (PIC) is a method based on statistical physics that ingeniously transforms the problem of finding optimal control strategies into solving expected trajectories\cite{williamsModelPredictivePath2015, williamsAggressiveDrivingModel2016, williamsInformationTheoreticMPC2017, nagabandiDeepDynamicsModels2019}. Additionally, the optimal trajectory computed by this method exhibits interesting symmetry-breaking phenomena\cite{kappenLinearTheoryControl2005, kappenPathIntegralsSymmetry2005}. In the PIC framework, optimal control actions are obtained through the weighted average of multiple sampled trajectories, with the update rule expressed as:

\begin{equation}
    \mu_t=\frac{\sum_{k=0}^{N-1}\left(e^{\tau R_k}\right)\left(a_t^{(k)}\right)}{\sum_{j=0}^{N-1} e^{\tau R_j}} \quad \forall t \in\{0, \ldots, H-1\},
    \label{eq:supp_pic_update}
\end{equation}
where \( \mu_t \) represents the optimal action mean at time step \( t \); \( N \) is the number of sampled trajectories; \( a_t^{(k)} \) is the action at time step \( t \) in the \( k \)-th trajectory; \( R_k \) is the cumulative reward value of the \( k \)-th trajectory; and \( \tau \) is a reward weight parameter used to balance exploration and exploitation. Formula \ref{eq:supp_pic_update} shows that the optimal action \( \mu_t \) is obtained by weighted averaging all sampled actions \( a_t^{(k)} \), with weights determined by the exponential function \( e^{\tau \cdot R_k} \); trajectories with higher cumulative rewards \( R_k \) have larger weights in calculating the optimal action.

However, this method may lead to the problem of local optima when the agent is in a local region of the parameter space. Since the algorithm relies on expectation calculations of finite-length trajectories, when system parameters are in a certain local region, the algorithm may converge to a locally optimal policy without exploring the globally optimal solution. Especially in physical system research, certain parameter regions may cause slow changes in system states, making it difficult for the agent to escape from the local state range.

\subsubsection{Reward-Error Adaptive Variance Control  (REAVC)}
\label{subsubsec:reavc}
To address the limitation of PIC methods potentially falling into local optima, the AMPPI algorithm proposes the Reward-Error Adaptive Variance Control (REAVC) strategy, inspired by the symmetry-breaking phenomenon of optimal trajectories in path integral control. The core idea of REAVC is to dynamically adjust the variance \( \Sigma_t \) of exploration actions to balance exploration and exploitation, and guide the algorithm to escape local optima. The adaptive update rule for variance \( \Sigma_t \) is as follows:
\begin{equation}
    \Sigma_t = \xi_t \Sigma_{t-1}, \quad \mathrm{where} \quad \xi_t = \mathrm{clip}\left(\exp(-\beta_{RE} \Delta r) - \eta_1 e_{\mathrm{inst}} - \eta_2 e_{\mathrm{hist}}, \varepsilon_{\min}, \varepsilon_{\max}\right)
    \label{eq:supp_reavc_variance_update}
\end{equation}    
where \( \Delta r = r_t - r_{t-1} \) is the reward change, measuring the magnitude of change in the current iteration reward \( r_t \) relative to the previous iteration reward \( r_{t-1} \); \( e_{\mathrm{inst}} = r_t - \alpha_{RE} \cdot r_{\mathrm{best}} \) is the instantaneous error term, representing the gap between the current reward \( r_t \) and the target reward level \( \alpha_{RE} \cdot r_{\mathrm{best}} \), with \( r_{\mathrm{best}} \) being the historical best exponentially weighted average reward value, used to set the benchmark for target rewards; \( e_{\mathrm{hist}} =  r_{t} - \alpha_{RE} \cdot r_{\mathrm{best}} \) is the historical error term, representing the deviation between the current exponentially weighted average reward \( r_{t} \) and the target reward level \( \alpha_{RE} \cdot r_{\mathrm{best}} \), where \( r_{t} \) is the exponentially weighted average reward at current time \( t \), updated iteratively through \( r_{t} = \lambda \cdot r_{t-1} + (1 - \lambda) r_t \), with \( \lambda \) being a forgetting factor that controls the weight of historical information. \( r_{\mathrm{best}} \) is updated after each iteration; if the current \( r_{t} \) is greater than \( r_{\mathrm{best}} \), then it is updated to \( r_{\mathrm{best}} = r_{t} \).

The parameters \( \alpha_{RE} \), \( \beta_{RE} \), \( \eta_1 \), and \( \eta_2 \) in formula \ref{eq:supp_reavc_variance_update} are key hyperparameters controlling the adaptive variance adjustment strategy, respectively controlling the influence of reward change, instantaneous error, and historical error on \( \Sigma_t \). The REAVC strategy implements a dynamic control mechanism for exploration variance by adjusting these parameters: when the reward value increases, the system reduces \( \Sigma_t \) for fine parameter adjustment; when the reward value decreases, the system increases \( \Sigma_t \) to encourage broader parameter space exploration. This adaptive adjustment mechanism is similar to temperature control in simulated annealing algorithms, helping the algorithm escape local optima and accelerate convergence to the global optimum.

To prevent the sampling variance \(\Sigma_t \) from becoming too small and causing action exploration stagnation, we introduce an adaptive variance lower bound adjustment mechanism. When all sampled actions have positive rewards, we dynamically adjust the variance based on reward distribution characteristics:

\begin{equation}
    \Sigma_t =
    \left\{
      \begin{array}{ll}
        \Sigma_t \cdot \left(\frac{r^* - \bar{r}}{\max(r) - \bar{r}}\right), & \mbox{if $\Sigma_t < \varsigma_{\mathrm{\min}}$ and $\min(r) > 0$} \\
        \Sigma_t, & \mbox{otherwise} 
      \end{array}
    \right. 
    \label{eq:adaptive_variance}
\end{equation}

where \(r = \{r^{(k)}\}_{k=1}^N \) represents the N sets of rewards sampled from the dynamic model \(f_\theta\), \(\bar{r} \) is the mean of these rewards, and \(\max(r) \) is the maximum reward value. \(r^* \) is the ideal reward value. By controlling the exploration covariance through \(\frac{r^* - \bar{r}}{\max(r) - \bar{r}}\), actions can better cover the target range.

\subsubsection{AMPPI: Integration of REAVC and PIC}
\label{subsubsec:amppi_integration}
The AMPPI algorithm innovatively integrates the advantages of the Reward-Error Adaptive Variance Control (REAVC) strategy and the Path Integral Control (PIC) method. Its core idea is to use REAVC to dynamically adjust the variance \( \Sigma_t \) of action sampling in the PIC method, thereby overcoming the limitation of PIC easily falling into local optima, and enhancing the efficiency and precision of parameter exploration in complex physical environments.

In each control cycle, the AMPPI algorithm utilizes the action sampling variance \( \Sigma_t \) calculated by the REAVC strategy from the previous control cycle. Based on this variance, multiple sets of action sequences are sampled from the Gaussian distribution \( \mathcal{N}(\mu, \Sigma_t) \). Subsequently, the dynamic model is used to predict the performance of these action sequences in the environment and calculate cumulative reward values. Finally, the optimal action \( \mu_t \) is calculated through the PIC update rule (formula \ref{eq:supp_pic_update}) and applied to the physical system. After the system executes the action and receives environmental feedback, the REAVC strategy calculates and updates the action sampling variance \( \Sigma_t \) based on the reward and error information from this iteration, preparing for action sampling in the next control cycle.

Having established the theoretical foundation and implementation details of the auxiliary search mechanism, we now present the complete AMPPI algorithm that integrates all components discussed in this section. The pseudocode below shows how the auxiliary reward mechanism is incorporated into the overall optimization framework:

\begin{algorithm}[H]
    \caption{Adaptive Model Predictive Path Integral (AMPPI)}
    \label{alg:amppi}
    \begin{algorithmic}[1]
    \footnotesize
    \Input{Number of exploration episodes \(N_{\exp}\), maximum exploration episode length \(T_{\exp}\), number of exploitation episodes \(N_{exploit}\), maximum exploitation episode length \(T_{exploit}\), number of action sequences \(N\), reward weight parameter \(\tau\), initial action variance \(\sigma_0^2 I\), REAVC parameters \(\alpha_{RE}\), \(\beta_{RE}\), \(\eta_1\), \(\eta_2\), \(\varepsilon_{\min}\), \(\varepsilon_{\max}\), \(\lambda\), \(\varsigma_{\min}\), ideal reward \(r^*\)}
    \Output{Optimal policy for parameter discovery}
    \State Initialize replay buffer \(\mathcal{D} = \emptyset\)
    \State Initialize exploration policy \(\pi_0 \) (e.g., random policy)
    \State Initialize dynamic model \(f_\theta \) with random parameters
    \State Initialize historical best reward \(r_{\mathrm{best}} = 0.5\)
    \State Initialize exponential moving average reward \(r_{t} = 0\)
    \State Initialize action variance \(\Sigma_t = \sigma_0^2 I \) \Comment{Initialize variance before exploitation}
    
    \For{\(episode = 1 \) to \(N_{\mathrm{exp, ep}}\)} \Comment{Exploration phase}
        \State Reset RL environment
        \For{\(t = 0 \) to \(L_{\mathrm{exp, ep}}\)}
            \State Select action \(a_t \) based on exploration policy \(\pi_0\)
            \State Execute action \(a_t \) in the physics environment, observe \(s_{t+1} \) and \(r_t\)
            \State Store transition \((s_t, a_t, s_{t+1}) \) in \(\mathcal{D}\)
        \EndFor
        \State Train dynamic model \(f_\theta \) using data \(\mathcal{D}\)
    \EndFor
    
    \For{\(episode = 1 \) to \(N_{\mathrm{exploit, ep}}\)} \Comment{Exploitation phase}
    \State Reset RL environment
    \State Initialize action mean \(\mu = 0\)
    \For{\(t = 0 \) to \(L_{\mathrm{exploit, ep}}\)}

        \State Sample action sequences \(\{a^{(k)}_{t:t+H-1}\}_{k=0}^{N-1} \sim \mathcal{N}(\mu, \Sigma_t) \) \Comment{Use current variance \(\Sigma_t\)}
        \State Predict state sequences \(\{s^{(k)}_{t+1:t+H}\}_{k=0}^{N-1} \) using \(f_\theta\)    
        \State Compute rewards \(\{r^{(k)}\}_{k=0}^{N-1} \) for each action sequence*
        \State Compute discounted return \(R_k = \sum_{h=0}^{H-1} \gamma_d^h r^{(k)}_{t+h}\)
        \State Compute optimal action sequences \(\mu_{t:t+H-1} \) where \(\mu_t = \frac{\sum_{k=0}^{N-1} (e^{\tau \cdot R_k}) a^{(k)}_t}{\sum_{j=0}^{N-1} e^{\tau \cdot R_j}} \) \Comment{PIC update}
        \State Execute action \(a^*_t=\mu_t \) in the physics environment, observe \(s_{t+1} \) and \(r_t\)
        \State Store transition \((s_t, a^*_t, s_{t+1}) \) in \(\mathcal{D}\)
        
        \State Compute reward change \(\Delta r = r_t - r_{t-1} \) (if \(t>0\), else \(\Delta r = 0\))
        \State Compute instantaneous error \(e_{\mathrm{inst}} = r_t - \alpha_{RE} \cdot r_{\mathrm{best}}\)
        \State Compute historical error \(e_{\mathrm{hist}} =  r_{t} - \alpha_{RE} \cdot r_{\mathrm{best}}\)
        \State Compute \(\xi_t = \mathrm{clip}\left(\exp(-\beta_{RE} \Delta r) - \eta_1 e_{\mathrm{inst}} - \eta_2 e_{\mathrm{hist}}, \varepsilon_{\min}, \varepsilon_{\max}\right)\)
        \State Update action variance \(\Sigma_{t+1} = \xi_t \Sigma_{t} \) \Comment{REAVC basic update}
        
        \State Update \(\Sigma_{t+1} \) according to adaptive variance mechanism in Equation (\ref{eq:adaptive_variance}) \Comment{Prevents exploration stagnation when variance is too small}
        
        \State Update exponential moving average reward \(r_{t} = \lambda \cdot r_{t-1} + (1 - \lambda) r_t\)
        \If {\(r_{t} > r_{\mathrm{best}}\)}
            \State Update \(r_{\mathrm{best}} = r_{t}\)
        \EndIf
    \EndFor
    \State Train dynamic model \(f_\theta \) using data \(\mathcal{D} \) \Comment{Retrain dynamic model with new data}
    \EndFor
    \item[] \textit{* Reward values computed as \(r^{(k)} = R(\hat{s}_t, a^{(k)}_t)\) per Section \ref{subsubsec:task_rewards}.}

    \end{algorithmic}
\end{algorithm}

\section{Implementation Details}
\label{sec:implementation}

\subsection{Reward Function Design}

\subsubsection{Parameter-Specific Reward Function}
\label{subsubsec:param_rewards}

Based on the general formulation in Eq. \ref{eq:supp_general_reward}, we define specific reward functions corresponding to different critical parameters. For the critical temperature \(T_c\), we utilize the Binder cumulant intersection property through \(R(U) = e^{-d(\mathcal{N}(U))}\), where \(U_L = 1 - \frac{\langle M^4 \rangle}{3 \langle M^2 \rangle^2}\) exhibits characteristic intersection behavior at the critical point across different system sizes.

The critical exponent \(\beta\) is determined using the scaled magnetization reward function \(R(\tilde{M}) = e^{-d(\mathcal{N}(\tilde{M}))}\), where \(\tilde{M}_L = M_L L^{\beta/\nu}\) represents the finite-size scaled magnetization. Similarly, the critical exponent \(\gamma\) corresponds to the scaled magnetic susceptibility through \(R(\tilde{\chi}) = e^{-d(\mathcal{N}(\tilde{\chi}))}\), where \(\tilde{\chi}_L = \chi_L L^{-\gamma/\nu}\) is the appropriately scaled magnetic susceptibility. Both scaled observables exhibit data collapse when the correct critical exponents are used.

\subsubsection{Auxiliary Search Mechanism for Non-uniqueness}
\label{subsubsec:aux_nu_search}

\paragraph{Problem Statement} 
As detailed in the main text, the global reward function $R_{\mathrm{global}}$ based on finite-size scaling exhibits an inherent non-uniqueness problem when simultaneously determining all critical parameters. Specifically, the data collapse quality for scaled magnetization $\tilde{M}$ and magnetic susceptibility $\tilde{\chi}$ depends only on the exponent ratios $\beta/\nu$ and $\gamma/\nu$, rather than the absolute values of individual exponents. This fundamental limitation prevents the direct determination of the correlation length exponent $\nu$, which is essential for complete critical parameter characterization.

To address this challenge and provide an unambiguous learning signal for $\nu$, we introduce an auxiliary search mechanism that leverages the unique scaling properties of the Binder cumulant. This section presents the theoretical foundation and algorithmic implementation of this mechanism.

\paragraph{Theoretical Foundation}
The auxiliary mechanism is grounded in the finite-size scaling theory for the Binder cumulant, which exhibits the universal scaling form:
\begin{equation}
U_L(T) = f_U\left((T - T_c)L^{1/\nu}\right)
\end{equation}
where $f_U$ is a universal scaling function independent of system size $L$. This scaling relation has a crucial implication: for the correct parameter combination $(T_c, \nu)$, Binder cumulant values from different system sizes $L_i$ and $L_j$ will collapse onto the same point on the universal curve when their scaling variables are identical:
\begin{equation}
(T_i - T_c)L_i^{1/\nu} = (T_j - T_c)L_j^{1/\nu} \Rightarrow U_{L_i}(T_i) = U_{L_j}(T_j)
\end{equation}
This property provides a direct test for the correctness of $\nu$ values, forming the theoretical basis for our auxiliary search mechanism.

\paragraph{Algorithm Design}
The auxiliary mechanism evaluates whether the current $\nu$ estimate satisfies the scaling collapse condition for Binder cumulants across different system sizes.

\textbf{Reference Point and Equivalent Temperatures:} 
For a reference system size $L_j$ at temperature $T_j = T_c + \varepsilon$ (where $\varepsilon$ is a small perturbation), the equivalent temperature for system size $L_i$ is:
\begin{equation}
T_i = T_c + (T_j - T_c) \left(\frac{L_j}{L_i}\right)^{1/\nu}
\label{eq:supp_equivalent_temp}
\end{equation}

\textbf{Auxiliary Reward Construction:}
The mechanism uses the learned dynamics model $f_\theta$ to predict Binder cumulant values $U_{L_i}(T_i)$ at the equivalent temperatures. The auxiliary reward is constructed as:
\begin{equation}
    R_{\mathrm{aux}}(\nu) = \exp(-d(\mathcal{N}(\{U_{L_i}(T_i), U_{L_j}(T_j), \ldots\})))
\end{equation}
where $d(\mathcal{N}(\cdot))$ measures the normalized variance across system sizes. When the $\nu$ estimate is correct, Binder cumulant values should be identical across different system sizes, maximizing $R_{\mathrm{aux}}$.

\paragraph{Details}
The perturbation parameter $\varepsilon$ is adaptively determined by $f_{\theta}$ to balance signal detectability and theoretical validity: too small values provide weak learning signals, while too large values violate finite-size scaling assumptions. Binder cumulant predictions employ ensemble averaging consistent with the global model architecture, ensuring robust auxiliary reward computation.

\subsubsection{Task-Specific Reward Integration}
\label{subsubsec:task_rewards}

The reward functions \(R(\cdot)\) defined in previous sections are used to compute reward values during algorithm execution, where \(r_t = R\) in the implementation. For single-parameter optimization tasks, we directly use the corresponding parameter-specific reward function from Section \ref{subsubsec:param_rewards}. For multi-parameter optimization, we employ composite reward structures:

For simultaneous determination of \((T_c, \beta, \gamma)\) with fixed \(\nu=1\), we use:
\begin{equation}
    R_{\mathrm{global}} = \frac{1}{3}[R(U) + R(\tilde{M}) + R(\tilde{\chi})]
\end{equation}

For complete four-parameter optimization \((T_c, \beta, \gamma, \nu)\), we combine global and auxiliary rewards:
\begin{equation}
    R = \frac{1}{2}[R_{\mathrm{global}} + R_{\mathrm{aux}}(\nu)]
\end{equation}

A staged optimization strategy is employed: initially optimizing \(T_c\) until \(R(U) \geq \varepsilon_{T_{c}}\), then activating joint optimization of all parameters.

\subsection{Dynamic Model Architecture and Processing}
\label{subsec:dynamic_model_impl}

\subsubsection{RNN Input Specification and Multi-Parameter Handling}
\label{subsubsec:rnn_input_spec}

The RNN-based dynamic model predicts changes in physical observables to maintain learning sensitivity in phase regions where observables vary minimally with temperature. The model processes current observables \(O_t = \{M, \chi, U\}\) and temperature adjustment \(\Delta T\), learning the mapping:
\begin{equation}
O_{t+1} = O_t + \mathrm{Denormalize}(f_\theta(O_t, \Delta T))
\label{eq:rnn_observable_prediction}
\end{equation}
where \(f_\theta(O_t, \Delta T)\) predicts normalized changes in observables rather than absolute values.

This differential prediction approach is crucial for learning stability, particularly when the system is in a stable phase where observables remain nearly constant with temperature variations, making it difficult for the model to learn meaningful patterns from minimal signal changes. By predicting observable differences rather than absolute values, the RNN maintains learning sensitivity even in phase regions where temperature adjustments produce negligible observable variations that would otherwise provide insufficient training signal for effective learning.

For multi-parameter optimization, the complete state update process involves two steps: First, the RNN predicts observables at the adjusted temperature. Second, the critical parameter updates \(\beta_{t+1} = \beta_t + \Delta\beta\), \(\gamma_{t+1} = \gamma_t + \Delta\gamma\), \(\nu_{t+1} = \nu_t + \Delta\nu\) are applied, and the predicted observables are transformed using finite-size scaling with the updated parameters:
\begin{eqnarray}
\tilde{M}_{L,t+1} &=& M_L \cdot L^{\beta_{t+1}/\nu_{t+1}} \\
\tilde{\chi}_{L,t+1} &=& \chi_L \cdot L^{-\gamma_{t+1}/\nu_{t+1}}
\end{eqnarray}

The complete state update combines raw and scaled observables:
\begin{equation}
s_{t+1} = (O_{t+1}, \tilde{O}_{t+1})
\label{eq:state_composition}
\end{equation}
where \(\tilde{O}_{t+1} = \{\tilde{M}_{L,t+1}, \tilde{\chi}_{L,t+1}\}\).

\subsubsection{Global and Local Model Switching Mechanism}
\label{subsec:global_local_switching}

To balance robustness and precision in dynamic predictions, we employ a threshold-based switching mechanism between the global and local model ensembles, driven by the Binder cumulant reward \( R(U) \), which indicates proximity to the critical temperature. The interaction is governed by the weighted ensemble in Eq.~\ref{eq:ensemble_prediction_weighted}, where the dynamic coefficient \( \alpha_M \in [0,1] \) controls the relative contribution: when \( R(U) < 0.5 \), \( \alpha_M = 1 \) favors the global ensemble for reliable broad exploration; once \( R(U) \geq 0.5 \), \( \alpha_M = 0 \) shifts to the local ensemble for sensitive refinement near criticality. 

% Experimental Configuration
\section{Experimental Configuration}
\subsection{Ising model environment configuration}
\label{subsec:ising_params}
Table \ref{tab:ising_model_params} provides the basic configuration parameters for the Ising model environment in this study. These parameter settings apply to all experiments in the main text unless otherwise specified. In the "Finding Critical Temperature" section experiments, we used different lattice size combinations \((L_1, L_2) \in \{(16,32), (32,64), (64,128)\}\), with other parameters remaining unchanged. The experiments in the "Simultaneously Finding Critical Temperature and Critical Exponents" section of the main text directly adopted the parameter configuration in this table, using a two-dimensional square lattice structure with lattice sizes \(L=32 \) and \(L=64\). In the "Transfer Learning" section, we changed the lattice type from square lattice to triangular lattice while keeping other parameter settings unchanged.

\begin{table}[htbp]
    \centering
    \caption{Ising model environment configuration}
    \label{tab:ising_model_params}
    \begin{tabularx}{\textwidth}{@{}Xc@{}}
        \toprule
        \textbf{Parameter} & \textbf{Value} \\
        \midrule
        Exchange coupling constant (\(J\)) & 1 \\
        Lattice type & 2D square lattice \\
        Lattice sizes (\(L\)) & 32, 64 \\
        Magnetic field (\(h\)) & 0.0 \\
        Initial temperature (\(T_{\mathrm{init}}\)) & \(T_{\mathrm{init}} \sim \mathcal{U}(0, T_{\mathrm{max}}) \) \\
        Relaxation steps & 5000 \\
        Sampling steps & 5000 \\
        Monte Carlo algorithm & Wolff algorithm~\cite{wolffCollectiveMonteCarlo1989} \\
        \bottomrule
    \end{tabularx}
\end{table}

The exchange coupling constant \( J \) is fixed at 1, indicating ferromagnetic interaction. Experiments are conducted under two different lattice sizes \( L \) (32 and 64). The initial temperature \( T_{\mathrm{init}} \) is randomly sampled from a uniform distribution \(\mathcal{U}(0, T_{\mathrm{max}})\), where \(T_{\mathrm{max}}\) is a predetermined maximum temperature value. At each temperature point, the system first undergoes 5000 relaxation steps to reach equilibrium, followed by 5000 sampling steps for calculating physical quantities. We employ the Wolff algorithm\cite{wolffCollectiveMonteCarlo1989} for Monte Carlo simulations.

\subsection{Hyperparameter Settings}
\label{subsec:hparams}

This section details the hyperparameter settings used in the experiments, including model training configuration, CEM algorithm configuration, and AMPPI algorithm configuration.

\paragraph{Model Training Configuration}
\label{subsec:nn_training}

The training configuration for the dynamic model is shown in Table \ref{tab:dynamic_model_hyperparameters}. We employ a two-layer recurrent neural network (RNN) structure, with each layer containing 50 hidden units and using tanh as the activation function. The output layer uses the identity function as the activation function. The optimizer chosen is Adam with a learning rate of \(1 \times 10^{-4}\).

\begin{table}[htbp]
    \centering
    \caption{Dynamic model training hyperparameters}
    \label{tab:dynamic_model_hyperparameters}
    \begin{tabularx}{\textwidth}{@{}Xc@{}}
        \toprule
        \textbf{Parameter} & \textbf{Value} \\
        \midrule
        Network Type & RNN \\
        Hidden Layers & 2 \\
        Hidden Units per Layer & 50 \\
        Activation Function & tanh \\
        Output Activation Function & identity \\
        Optimizer & Adam \\
        Learning Rate & \(1 \times 10^{-4} \) \\
        Dropout Rate & 0.0 \\
        Batch Size (Train) & 5000 \\
        Batch Size (Validation) & 5000 \\
        Epochs & 100 - 50000 \\
        Early Stopping Patience (p) & 40 \\
        Early Stopping Threshold (\(\epsilon\)) & \(1 \times 10^{-4} \) \\
        \(N_g\) (Global) & 2 \\
        \(N_l\) (Local) & 3 \\
        New Data Percentage & 0.8 \\
        \bottomrule
    \end{tabularx}
\end{table}

\paragraph{CEM Configuration}
\label{subsubsec:cem_config}

The configuration parameters for the Cross Entropy Method (CEM) are shown in Table \ref{tab:cem_hyperparameters}. The maximum number of iterations is set to 500, with an elite sample ratio of 0.1 (corresponding to 10 samples).

\begin{table}[htbp]
    \centering
    \caption{CEM hyperparameters}
    \label{tab:cem_hyperparameters}
    \begin{tabularx}{\textwidth}{@{}Xc@{}}
        \toprule
        \textbf{Parameter} & \textbf{Value} \\
        \midrule
        Max Iterations & 500 \\
        Number of Elites & 10 \\
        \bottomrule
    \end{tabularx}
\end{table}
 
\paragraph{AMPPI Configuration}
\label{subsubsec:amppi_config}

The configuration parameters for the AMPPI algorithm are shown in Table \ref{tab:amppi_hyperparameters}. The prediction horizon \( H \) is set to 10, with 500 sampled action sequences. The parameters related to the REAVC strategy are also listed in this table.

\begin{table}[htbp]
    \centering
    \caption{AMPPI hyperparameters}
    \label{tab:amppi_hyperparameters}
    \begin{tabularx}{\textwidth}{@{}Xc@{}}
        \toprule
        \textbf{Parameter} & \textbf{Value} \\
        \midrule
        Horizon (\(H\)) & 10 \\
        Number of Action Sequences & 500 \\
        Discount factor (\(\gamma_d\)) & 0.95 \\
        Reward Weight Parameter (\(\tau\)) & 100.0* \\
        Initial Action Variance (\(\sigma_0\)) &  0.1 \\
        \midrule
        \multicolumn{2}{c}{\textbf{REAVC Parameters}} \\
        \midrule
        \(\alpha_{RE} \) & 0.9 \\
        \(\beta_{RE} \) & 0.2 \\
        \(\eta_1 \) &  0.2 \\
        \(\eta_2 \) & 0.1 \\
        \(\varepsilon_{\min} \) & 0.9 \\
        \(\varepsilon_{\max} \) & 1.1 \\
        \(\varepsilon_{T_{c}} \) & 0.9 \\
        \(\lambda_d \) & 0.95 \\
        \(\varsigma_{\min} \) & \(1 \times 10^{-2} \)  \\
        \midrule
        \multicolumn{2}{c}{\textbf{Trajectory Settings}} \\
        \midrule
        Number of Random Exploration Episodes (\(N_{\mathrm{exp, ep}}\)) & 20 \\
        Random Exploration Episode Length (\(L_{\mathrm{exp, ep}}\)) & 100 \\
        Total Number of Exploration Episodes (\(N_{\mathrm{exp}}\)) & \(N_{\mathrm{exp, ep}} \cdot L_{\mathrm{exp, ep}}\) \\
        Number of Exploitation Episodes (\(N_{\mathrm{exploit, ep}}\)) & 20 \\
        Exploitation Episode Length (\(L_{\mathrm{exploit, ep}}\)) & 200 \\
        \bottomrule
        \multicolumn{2}{l}{\footnotesize * Indicates parameters that were varied or not used in specific experiments.}
    \end{tabularx}
\end{table}

\subsection{Pyfssa Configuration}
\label{subsec:pyfssa_config}

To evaluate the performance difference between the AMPPI algorithm and traditional methods in critical parameter identification, we chose pyfssa as a benchmark comparative method. Pyfssa is a Python-based Finite-Size Scaling Analysis toolkit\cite{sorgePyfssa0762015} widely used for automatic analysis of system critical parameters.

The core principle of pyfssa is to automate the "data collapse" technique. According to finite-size scaling theory (Eq.~\ref{eq:finite_size_scaling}), when observable data from different system sizes $L$ are rescaled using the correct critical parameters ($T_c$ and critical exponents), they should collapse onto a single universal scaling function. Pyfssa quantifies the quality of this collapse by defining a cost function that measures the dispersion of the rescaled data points relative to a common master curve. By numerically minimizing this cost function (typically using the Nelder-Mead algorithm\cite{nelderSimplexMethodFunction1965}), the toolkit identifies the optimal set of critical parameters that produces the best data collapse\cite{houdayerLowtemperatureBehaviorTwodimensional2004}.

In this study, we uniformly sampled 100 temperature points in the reduced temperature range \(t = \frac{T - T_c}{T_c} \in [-0.2, 0.2]\) (corresponding to a temperature range of approximately \(0.8T_c \) to \(1.2T_c\)), and densely sampled 60 points in the range \(t \in [-0.1, 0.1] \) (corresponding to a temperature range of \(0.9T_c \) to \(1.1T_c\)) to ensure higher precision near the critical point.

For each temperature point, we used the same Monte Carlo simulation parameters as the AMPPI algorithm (see Table \ref{tab:ising_model_params}) to generate physical observable data for the Ising model. To maximize the precision of pyfssa, we sampled 80 sets of observable data at the same temperature range for different environment scales \(L\) and calculated the mean and standard deviation. Subsequently, automatic scaling analysis was performed to obtain critical parameters. Specifically, the critical temperature \(T_c\) was first obtained through the Binder cumulant, with the critical exponent \(\mu = 1\) fixed, then scaling analysis was performed on magnetization \(M\) and magnetic susceptibility \(\chi\) to obtain critical exponents \(\beta\) and \(\gamma\).

To ensure the stability and reliability of optimization results, we set initial value ranges for each critical parameter and uniformly sampled 100 parameter starting points within these ranges. The final critical parameters were obtained by minimizing the cost function for data collapse quality\cite{houdayerLowtemperatureBehaviorTwodimensional2004}. The initial value ranges for each critical parameter were:
\(T \in [0.01, 10], \quad \beta \in [-10, 10], \quad \gamma \in [-10, 10]\)

% Extended Analysis
\section{Experimental Analysis}
\label{sec:analysis}
\subsection{Model Performance}
\label{subsec:model_perf}
% Training curves, prediction accuracy

To evaluate the performance of the dynamic model in predicting the evolution of the Ising model environment, we conducted a detailed analysis of the model's training process and prediction accuracy. As described in \ref{subsubsec:model_learning}, we constructed a dynamic model using recurrent neural networks (RNNs) and trained it using the loss function defined in equation \ref{eq:loss_function_normalized}.

\begin{figure}[htbp]
    \centering
    \includegraphics[width=\linewidth]{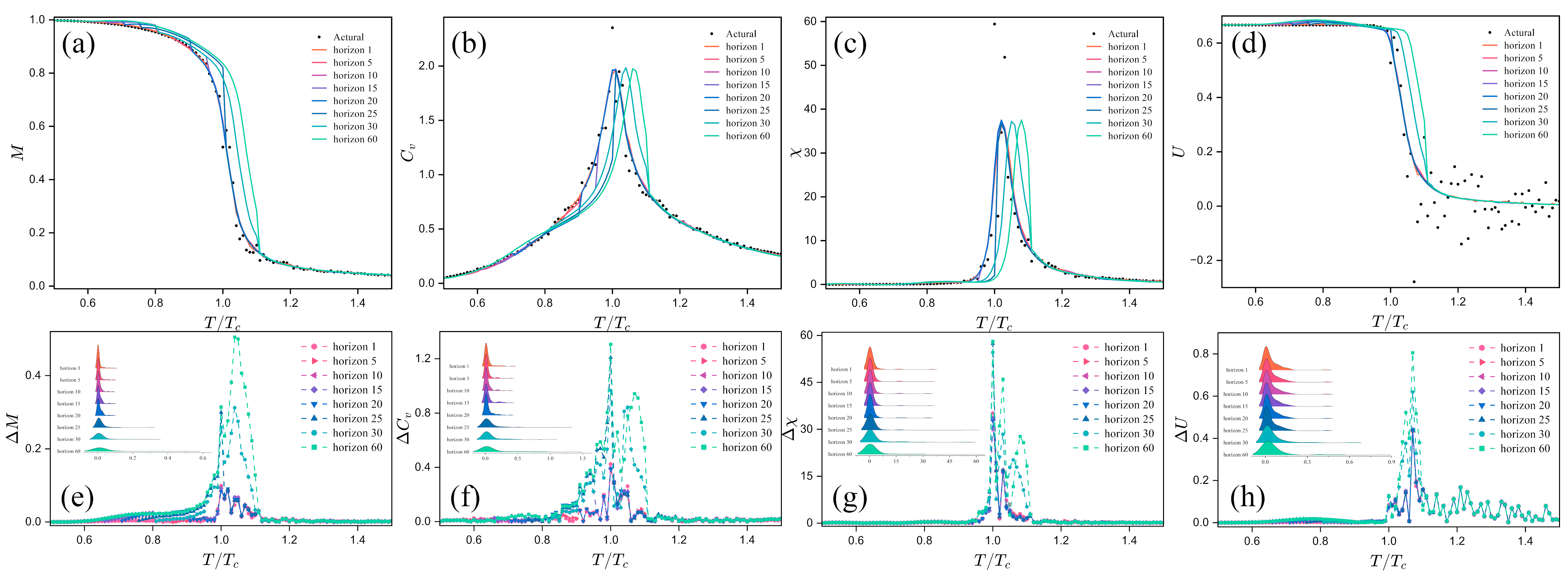}
    \captionsetup{justification=raggedright, singlelinecheck=false}
    \caption{Analysis of the dynamic model's prediction performance for Ising model observables. (a)-(d) show the prediction results for physical quantities at different prediction horizons for system size \(L=50\): (a) magnetization \(M\), (b) specific heat \(C_v\), (c) magnetic susceptibility \(\chi\), and (d) Binder cumulant \(U\). Solid lines represent model predictions, and black dots represent actual observations. (e)-(h) analyze the evolution of prediction errors with prediction horizon: the main plots show absolute error changes at different temperatures, and the insets show the statistical distribution of errors.}
    \label{fig:state_prediction}
\end{figure}

Figure \ref{fig:state_prediction} shows the dynamic model's prediction results for Ising model observables (magnetization, specific heat, magnetic susceptibility, and Binder cumulant) at different prediction horizons. The results indicate that prediction errors gradually increase with the prediction horizon, which is expected. However, even at longer prediction horizons, the model can still capture the overall trend of the observables. Particularly when \( H \leq 20 \), predictions at different horizons are already difficult to distinguish (Figure \ref{fig:state_prediction} (e)-(h)).

\begin{figure}[htbp]
    \centering
    \includegraphics[width=\linewidth]{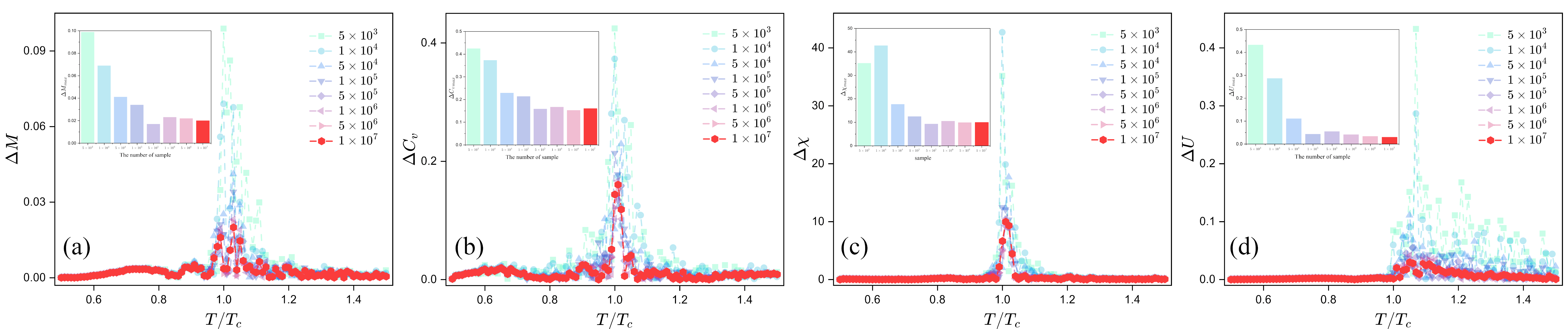}
    \captionsetup{justification=raggedright, singlelinecheck=false}
    \caption{Error analysis of dynamic model predictions compared to observables with different sampling counts. The main plot shows the difference between model predictions and actual observations obtained with different sampling counts (\(5 \times 10^3 \) to \(1 \times 10^7\)), with the x-axis representing reduced temperature \(T/T_c\). The model was trained using data from \(5 \times 10^3 \) sampling iterations. The inset shows the trend of maximum prediction error with sampling count.}
    \label{fig:diff_iter_actural_vs_pre}
\end{figure}
 
Interestingly, we found that even a dynamic model trained with lower precision sampling data (\(5 \times 10^3\) iterations) can make relatively accurate predictions for observables under high-precision sampling (\(5 \times 10^5\) iterations and above) (Figure \ref{fig:diff_iter_actural_vs_pre}). Especially when \(T \geq T_c\), the dynamic model can accurately predict physical quantities under high-precision sampling. This indicates that the dynamic model can effectively infer system dynamic behavior from observables with larger perturbations, demonstrating strong robustness.

Overall, the dynamic model shows good performance in predicting the evolution of the Ising model environment. Even with a longer prediction horizon, the model can capture the overall trend of observables and exhibits robustness to noise in the environment.

\subsection{Algorithmic Trajectory}
\label{subsec:param_traj}
% Parameter space visualization

To more intuitively understand the search behavior of the AMPPI algorithm in parameter space, we conducted a visualization analysis of the algorithm's trajectory during the process of finding critical temperature and critical exponents.

\begin{figure}[htbp]
    \centering
    \includegraphics[width=\linewidth]{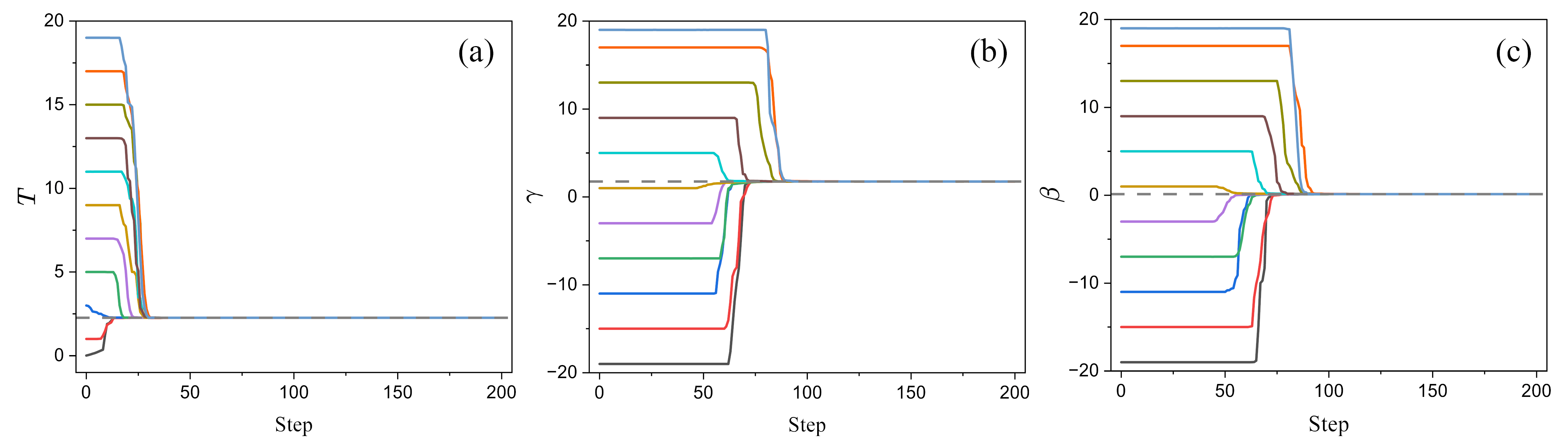}
    \captionsetup{justification=raggedright, singlelinecheck=false}
    \caption{Search trajectory of the AMPPI algorithm in parameter space. Dashed lines represent theoretical values for each parameter.}
    \label{fig:parameter_trajectories}
\end{figure}

For critical temperature \(T_c \) and critical exponents \(\beta \) and \(\gamma\), the algorithm's search trajectories all exhibit a significant feature: in the early stage of the search, although parameter values fluctuate slightly and seem to "stagnate," during this phase, the action sampling variance within the algorithm is gradually increasing. When the sampling variance reaches a certain critical value, the parameter trajectory suddenly exhibits clear directionality, rapidly moving toward the target value region. This trajectory behavior indicates an essential characteristic of the algorithm, which we will analyze more deeply in section \ref{subsec:explore_analyze}.

\subsection{Algorithm Behavior Analysis}
\label{subsec:explore_analyze}
% Variance adaptation dynamics

\begin{figure}[htbp]
    \centering
    \includegraphics[width=0.8\linewidth]{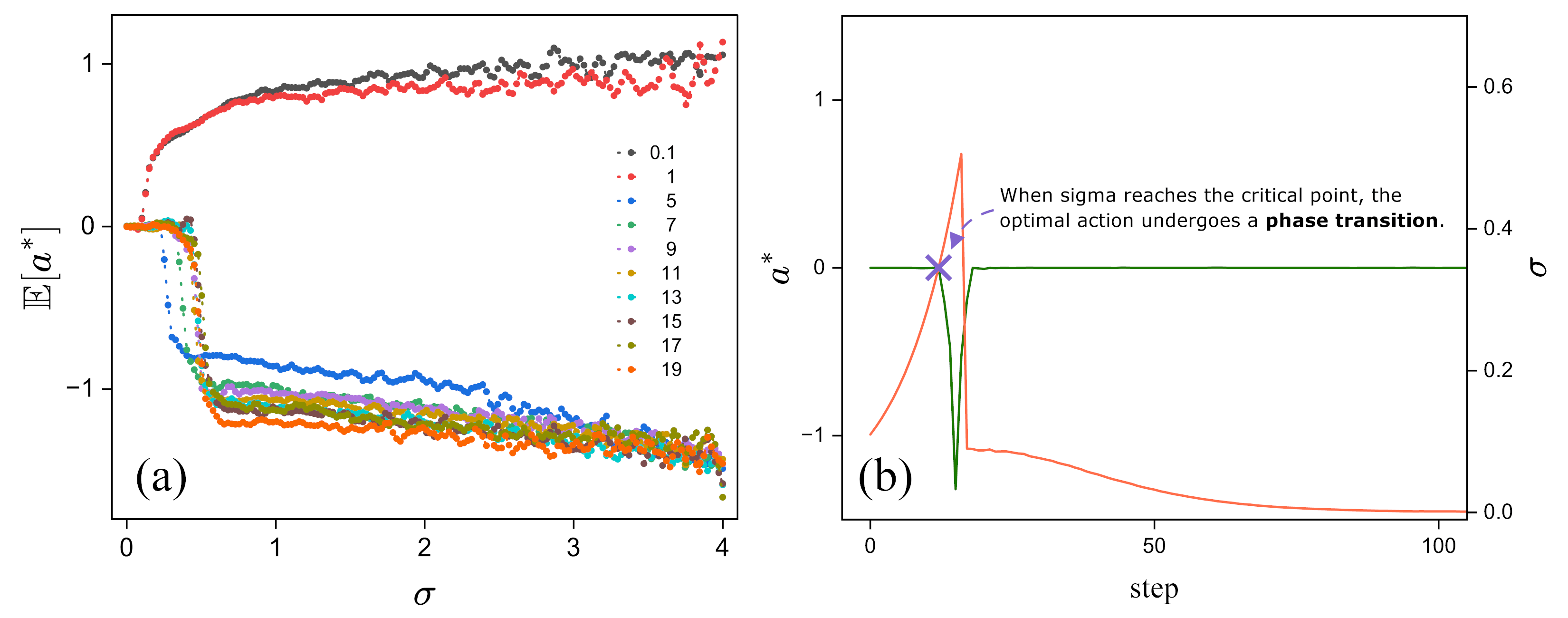}
    \captionsetup{justification=raggedright, singlelinecheck=false}
    \caption{Symmetry breaking phenomenon in action space. (a) Shows the evolution of the expected value of optimal actions \(\mathbb{E}[a^*] \) with exploration variance \(\sigma \) at different initial temperatures \(T\). (b) Shows the evolution of optimal action \(a^* \) and exploration variance \(\sigma \) with action steps at initial temperature position \(T=5\).}
    \label{fig:phase_transition}
\end{figure}

To gain a deeper understanding of the behavioral mechanism of the AMPPI algorithm, we conducted a detailed analysis of its key component—the Reward-Error Adaptive Variance Control (REAVC) strategy. Figure \ref{fig:phase_transition} shows the phase transition phenomenon observed in action space, which bears profound similarities to symmetry breaking in physical systems.

The action behavior exhibits remarkable regularity: the expected value of optimal actions \(\mathbb{E}[a^*]\) undergoes qualitative transitions with changes in exploration variance \(\sigma\) under different initial temperature conditions. When the variance is small, optimal actions fluctuate around zero, exhibiting a "disordered" state; but when the exploration variance exceeds a specific critical threshold, optimal actions suddenly show a clear directional preference, and the system undergoes a transition from "disorder" to "order." This characteristic demonstrates a precise correspondence with phase transition processes in physical systems, with exploration variance \(\sigma\) functionally equivalent to the temperature parameter in thermodynamics, and the directionality of optimal actions corresponding to the order parameter in phase transition theory. Notably, the critical threshold of transition variance exhibits a positive correlation with the distance from the initial state to the target state—the farther the initial position is from the target position, the larger the critical variance value needed to trigger the action transition, suggesting an inherent phase transition mechanism in the system.

Figure \ref{fig:phase_transition}(b) provides this phase transition process: in the initial stage, the REAVC strategy gradually increases the action sampling variance, and at step 13, when variance \(\sigma = 0.37975\), the optimal action \(a^*\) suddenly jumps from -0.00125 to 0.2, marking a qualitative transition within the algorithm.

This "phase transition" behavior endows the AMPPI algorithm with an "intuition-like" capability. By introducing varying "temperature," the algorithm can adaptively adjust the exploration range in parameter space and break the symmetry of actions when the algorithm is in a local optimum, quickly finding the "shortcut" to target parameters. This physically inspired design enables the AMPPI algorithm to exhibit excellent exploration efficiency and robustness in complex environments.

% Implementation Details

\subsection{Detailed Results for Multi-parameter search}
\label{subsec:detailed_multi_param_results}

To more deeply analyze the performance of the AMPPI algorithm in simultaneously finding critical temperature and critical exponents, we provide detailed results at different iteration counts in Table \ref{tab:detailed_multi_param_search}. The table shows the AMPPI algorithm's estimated values for critical temperature \(T_c\) and critical exponents \((\beta, \gamma)\) at 1 iteration, 20 iterations, 100 iterations, and 200 iterations, under different system scales \(L=(16,32), (32, 64), (64, 128)\), compared with the best results from the pyfssa algorithm. The table shows that as the number of iterations increases, the AMPPI algorithm's estimation precision for critical parameters continuously improves, and errors gradually decrease. Even after only a few iterations, the AMPPI algorithm's performance is significantly better than the traditional pyfssa algorithm.

\begin{table}[htbp]
    \centering
    \resizebox{\textwidth}{!}{
    \begin{tabular}{ccccccc}
        \toprule
        \multirow{2}{*}{Environment Size} & \multirow{2}{*}{Parameter} &  \multicolumn{4}{c}{Iterations}& \multirow{2}{*}{pyfssa(best)}\\
        \cmidrule(lr){3-6}
        & &  1 Iteration&20 Iterations& 100 Iterations& 200 Iterations& \\
        \midrule
        \multirow{3}{*}{\(L=(16,32)\)}
        & \(T_c \) &  2.2704 \(\pm\) 0.0224&2.2684 \(\pm\) 0.0009& 2.2680 \(\pm\) 0.0007& 2.2681 \(\pm\) 0.0006& \(2.2574\pm 0.0055\)\\
        & \(\beta \) &  0.1307 \(\pm\) 0.1076&0.1221 \(\pm\) 0.0021& 0.1212 \(\pm\) 0.0013& 0.1223 \(\pm\) 0.0018& \(0.0943\pm 0.006\)\\
        & \(\gamma \) &  1.7505 \(\pm\) 0.1944&1.7470 \(\pm\) 0.0073& 1.7464 \(\pm\) 0.0042& 1.7518 \(\pm\) 0.0076& \(1.7578\pm 0.1247\)\\
        \midrule
        \multirow{3}{*}{\(L=(32, 64)\)}
        & \(T_c \) &  2.2711 \(\pm\) 0.0089&2.2693 \(\pm\) 0.0007& 2.2690 \(\pm\) 0.0003& 2.2690 \(\pm\) 0.0003& \(2.2573\pm 0.0052\)\\
        & \(\beta \) &  -0.0418 \(\pm\) 0.0596&0.1244 \(\pm\) 0.0017& 0.1235 \(\pm\) 0.0008& 0.1247 \(\pm\) 0.0009& \(0.0930\pm 0.0042\)\\
        & \(\gamma \) &  1.7204 \(\pm\) 0.2351&1.7477 \(\pm\) 0.0107& 1.7501 \(\pm\) 0.0030& 1.7490 \(\pm\) 0.0043& \(1.8228\pm 0.1472\)\\
        \midrule
        \multirow{3}{*}{\(L=(64, 128)\)}
        & \(T_c \) &  2.2687 \(\pm\) 0.0356&2.2692 \(\pm\) 0.0006& 2.2692 \(\pm\) 0.0003& 2.2691 \(\pm\) 0.0002& \(2.2622\pm 0.0106\)\\
        & \(\beta \) &  0.1222 \(\pm\) 0.1116&0.1163 \(\pm\) 0.0055& 0.1244 \(\pm\) 0.0017& 0.1251 \(\pm\) 0.0016& \(0.096\pm 0.005\)\\
        & \(\gamma \) &  1.6145 \(\pm\) 0.1677&1.7584 \(\pm\) 0.0277& 1.7503 \(\pm\) 0.0121& 1.7508 \(\pm\) 0.0078& \(1.8469\pm 0.1265\)\\
        \bottomrule
    \end{tabular}
    }
    \captionsetup{justification=raggedright, singlelinecheck=false}
    \caption{Detailed Performance of AMPPI for Simultaneous Critical Parameter Search. The table presents the estimated values of critical temperature (\(T_c\)) and critical exponents (\(\beta\), \(\gamma\)) for the 2D Ising model at different iteration counts and system sizes, comparing them with the best results from the pyfssa algorithm.}
    \label{tab:detailed_multi_param_search}
\end{table}

\newpage

\subsection{Detailed Results for Transfer Learning}
\label{subsec:detailed_transfer_learning_results}

To more thoroughly evaluate the performance of the AMPPI algorithm in transfer learning tasks, we provide detailed experimental results in Table \ref{tab:detailed_transfer_learning}. The table compares the performance of direct transfer method, sample fine-tuning method (showing results after 1 iteration, 20 iterations, and 100 iterations of fine-tuning), and the pyfssa algorithm in finding critical temperature \(T_c\) and critical exponents \((\beta, \gamma)\) in the triangular lattice Ising model under different system scales \(L=(16,32), (32, 64), (64, 128)\), compared with the best results from the pyfssa algorithm. The data clearly demonstrates the effectiveness of the sample fine-tuning method in enhancing transfer learning performance, and the AMPPI algorithm still outperforms the traditional pyfssa algorithm in transfer learning scenarios.

\begin{table}[htbp]
    \centering
    \resizebox{\textwidth}{!}{
    \begin{tabular}{ccccccc}
        \toprule
        \multirow{2}{*}{Environment Size} & \multirow{2}{*}{Parameter} & \multirow{2}{*}{Direct Transfer} & \multicolumn{3}{c}{Sample Fine-tuning} & \multirow{2}{*}{pyfssa(best)}\\
        \cmidrule(lr){4-6}
        & & & 1st Iteration& 5th Iteration& 20th Iteration& \\
        \midrule
        \multirow{3}{*}{\(L=(16,32)\)} & \(T_c \) & 3.6050 \(\pm\) 0.0480& 3.6390 \(\pm\) 0.0073& 3.6396 \(\pm\) 0.0022
        & 3.6407 \(\pm\) 0.0008
        & \(3.6365 \pm 0.0141\)\\
                & \(\beta \) & 0.0880 \(\pm\) 0.0197& 0.1141 \(\pm\) 0.0161& 0.1250 \(\pm\) 0.0046
        & 0.1232 \(\pm\) 0.0014
        & \(-0.0951 \pm 0.0066\)\\
                & \(\gamma \) & 1.6112 \(\pm\) 0.1472& 1.7330 \(\pm\) 0.0623& 1.7599 \(\pm\) 0.0218& 1.7470 \(\pm\) 0.0041& \(1.8192\pm 0.1101\)\\
                \midrule
                \multirow{3}{*}{\(L=(32, 64)\)} & \(T_c \) & 3.6661 \(\pm\) 0.0510& 3.6394 \(\pm\) 0.0089& 3.6383 \(\pm\) 0.0019
        & 3.6401 \(\pm\) 0.0009
        & \(3.6391 \pm 0.0038\)\\
                & \(\beta \) & 0.1348 \(\pm\) 0.1602& 0.1146 \(\pm\) 0.0468& 0.1224 \(\pm\) 0.0073
        & 0.1246 \(\pm\) 0.0013
        & \(0.0964 \pm 0.0067\)\\
                & \(\gamma \) & 1.7507 \(\pm\) 0.1715& 1.7404 \(\pm\) 0.1406& 1.7501 \(\pm\) 0.0157& 1.7508 \(\pm\) 0.0093& \(1.8468 \pm 0.1122\)\\
                \midrule
                \multirow{3}{*}{\(L=(64, 128)\)} & \(T_c \) & 3.6382 \(\pm\) 0.0044& 3.6437 \(\pm\) 0.0200
        & 3.6402 \(\pm\) 0.0011
        & 3.6407 \(\pm\) 0.0004
        & \(3.6401 \pm 0.0023\)\\
                & \(\beta \) & 0.0350 \(\pm\) 0.0695& 0.1172 \(\pm\) 0.0616
        & 0.1232 \(\pm\) 0.0270
        & 0.1244 \(\pm\) 0.0011
        & \(0.0939 \pm 0.0064\)\\
        & \(\gamma \) & 1.3143 \(\pm\) 0.2395& 1.5112 \(\pm\) 0.1384& 1.7201 \(\pm\) 0.0554& 1.7477 \(\pm\) 0.0052& \(1.8799 \pm 0.1112\)\\
        \bottomrule
    \end{tabular}
    }
    \captionsetup{justification=raggedright, singlelinecheck=false}
    \caption{Detailed Performance of Transfer Learning for Critical Parameter Search in Triangular Lattice Ising Model. The table compares the performance of direct transfer and sample fine-tuning methods for estimating critical temperature (\(T_c\)) and critical exponents (\(\beta\), \(\gamma\)) in the triangular lattice Ising model at different system sizes, along with the best results from the pyfssa algorithm.}
    \label{tab:detailed_transfer_learning}
\end{table}
\newpage

\section{Implementation Details}
\label{sec:impl}
% \subsection{Codebase Architecture}
% \label{subsec:code_struct}
% Module dependency diagram

% \subsection{Computational Resources}
% \label{subsec:compute_res}
% GPU/CPU utilization metrics
All experiments in this study were conducted on a server equipped with two Intel(R) Xeon(R) Gold 6230 CPUs @ 2.10GHz, each containing 20 physical cores (a total of 40 physical cores, 80 logical cores), with 125GiB server memory. All computational tasks were completed on CPUs without GPU acceleration. The dynamic model was constructed using the PyTorch framework, while the Ising model simulation was written in Fortran and compiled. The average action computation time for the AMPPI algorithm was 1.25 seconds, and for the CEM algorithm was 5.98 seconds. For the Ising model, the average running times for simulations using the Wolff algorithm across different lattice types and scales were as follows: square lattice \(L=32 \) was 0.30 seconds, \(L=64 \) was 3.58 seconds, \(L=128 \) was 7.83 seconds; triangular lattice \(L=16 \) was 0.52 seconds, \(L=32 \) was 1.74 seconds, \(L=64 \) was 5.93 seconds, \(L=128 \) was 21.12 seconds. To fully utilize computational resources, we employed optimization measures such as multi-process parallelization, code optimization, and Fortran acceleration.

\vskip 0.2in

\end{document}